\documentclass[a4paper,twocolumn,conference]{IEEEtran}
\usepackage[T1]{fontenc}
\usepackage[latin9]{inputenc}
\usepackage{array}
\usepackage{verbatim}
\usepackage{algorithm2e}
\usepackage{amsmath}
\usepackage{amssymb}
\usepackage{graphicx}
\usepackage[unicode=true,
 bookmarks=true,bookmarksnumbered=true,bookmarksopen=true,bookmarksopenlevel=1,
 breaklinks=false,pdfborder={0 0 0},pdfborderstyle={},backref=false,colorlinks=false]
 {hyperref}
\hypersetup{pdftitle={Your Title},
 pdfauthor={Your Name},
 pdfpagelayout=OneColumn, pdfnewwindow=true, pdfstartview=XYZ, plainpages=false}
\usepackage{breakurl}

\makeatletter

\providecommand{\tabularnewline}{\\}

\usepackage[caption=false,font=footnotesize]{subfig}

\LinesNumbered
\makeatother

\begin{document}

\title{Defense Methods Against Adversarial Examples for Recurrent Neural
Networks}

\author{\IEEEauthorblockN{Ishai~Rosenberg}\IEEEauthorblockA{Ben Gurion University of the Negev}\and \IEEEauthorblockN{Asaf~Shabtai}\IEEEauthorblockA{Ben Gurion University of the Negev}\and \IEEEauthorblockN{Yuval~Elovici}\IEEEauthorblockA{Ben Gurion University of the Negev}\and \IEEEauthorblockN{Lior~Rokach}\IEEEauthorblockA{Ben Gurion University of the Negev}}
\maketitle
\begin{abstract}
Adversarial examples are known to mislead deep learning models to
incorrectly classify them, even in domains where such models achieve
state-of-the-art performance. \\
Until recently, research on both attack and defense methods focused
on image recognition, primarily using convolutional neural networks
(CNNs). In recent years, adversarial example generation methods for
recurrent neural networks (RNNs) have been published, demonstrating
that RNN classifiers are also vulnerable to such attacks. \\
In this paper, we present a novel defense method, termed sequence
squeezing, to make RNN classifiers more robust against such attacks.
Our method differs from previous defense methods which were designed
\emph{only} for non-sequence based models. We also implement four
additional RNN defense methods inspired by recently published CNN
defense methods. \\
We evaluate our methods against state-of-the-art attacks in the cyber
security domain where real adversaries (malware developers) exist,
but our methods can be applied against other discrete sequence based
adversarial attacks, e.g., in the NLP domain. Using our methods we
were able to decrease the effectiveness of such attack from 99.9\%
to 15\%.
\end{abstract}

\section{Introduction}

The growing use of deep learning in fields like computer vision and
natural language processing (NLP) \cite{DBLP:journals/nature/LeCunBH15},
has been accompanied by increased interest in the domain of adversarial
learning, that is, attacking and defending deep learning models algorithmically.
Of special interest are adversarial examples, which are samples slightly
modified in order to be misclassified by the attacked classifier. 

Most of the research in deep adversarial learning has focused mainly
on convolutional neural networks (CNNs) commonly used in the computer
vision domain, and more specially, in the image recognition domain
\cite{DBLP:journals/access/AkhtarM18}. However, in recent years,
more and more adversarial example generation methods have been presented
in the NLP domain in order to bypass recurrent neural network (RNN)
classifiers, e.g., sentiment analysis classifiers \cite{JiDeepWordBug18}. 

Adversarial attacks have also been used in the cyber security domain.
This domain raises special interest, because it involves adversaries:
malware developers who want to evade next generation machine and deep
learning based classifiers. Such attacks have already been executed
against static analysis non sequential deep neural networks \cite{CylanceAdvAttack}. 

The threat of such attacks makes adversarial attacks against RNN classifiers
an interesting and important real-life use case, especially in the
cyber security domain. For this reason, we focus this paper on RNN
defense methods in the cyber security domain and not, for instance,
on the more heavily researched NLP domain. Thus, we only focus on
discrete sequence input (e.g., discrete API call type sequences).
Evaluating the methods presented in this paper in other domains where
the input is a sequence, such as NLP, will be a part of our future
work.

The most prominent use case of RNN classifiers in the cyber security
domain is analyzing API calls of a running process as features \cite{DBLP:journals/corr/abs-1806-10741,Huang2016,Kolosnjaji2016,Pascanu2015}.
API call based RNN classifiers have superior performance in comparison
to their CNN equivalents \cite{Athiwaratkun2017,DBLP:conf/raid/RosenbergSRE18}.
This results from the RNN classifiers' ability to leverage the context
of the malicious API calls, using their hidden state as context memory.
In contrast, CNN classifiers can only use adjacent API calls as a
context, due to their spatial locality, which is less relevant when
analyzing long API call traces.

Attacks against API call based RNN classifiers have already been published
\cite{DBLP:journals/corr/abs-1804-08778,DBLP:conf/raid/RosenbergSRE18}.
Thus, in-order to use such classifiers in an adversarial setting,
we need adversarial defense methods that works with API call based
RNN classifiers, which is the focus of this paper.

One might claim that a defense method that does not block 100\% of
the adversarial examples is insufficient in certain domains, such
as the cyber security domain. We consider two cases. The first is
an attacker that wants to infect a target with a \emph{specific},
perhaps specially crafted, malware (e.g., the WannaCry ransomware,
Cosmic Duke APT, etc.). If this adversarial example cannot evade the
malware classifier, the attacker must invest a lot of time generating
another malware to use against the target host(s). Thus, defense methods
that block 85\% of the attacks, as presented in this paper, have a
significant value. In the second case, the attacker holds an arsenal
of malware, and he/she would like to have any of then successfully
bypassing our detection mechanism. In this case, 100\% detection rate
is needed. However, we consider this use case to be less realistic.

The contributions of our paper are as follows:

1) We present \emph{sequence squeezing}, a novel defense method that
reduces the adversarial space and limits the possibility of generating
adversarial input sequences. This reduction is performed without modifying
the classifier. The reduction uses a dedicated sequence transformation,
since image based input transformations (such as reducing the image
depth) cannot be applied to sequence inputs.

2) We present four additional defense methods, inspired by CNN defenses: 

(i) A method that uses adversarial examples' statistical properties,
leveraging the correlation between items inside the sequence. This
has not been done in previous research which uses non-sequence inputs.

(ii) A method that uses several different subsequences inside the
input sequence as training sets for several models used in an ensemble.This
prevents adversarial examples with localized modifications from fooling
the entire ensemble. This method cannot be used on non-sequence input. 

(iii) A method that trains a generative adversarial network (GAN)
on the training set and uses the output that is closest to the original
input instead of the input sequence itself. This is done in order
to remove the adversarial perturbations before classification. GANs
used for image generation cannot be used here, because the discrete
outputs from the generative model (e.g., API call type sequences)
make it difficult to pass the gradient update from the discriminative
model to the generative model. 

(iv) A method that classifies the input sequence's nearest neighbor
in the training set instead of the input sequence itself. This is
done in order to remove the adversarial perturbations before classification.

To the best of our knowledge, there is no paper addressing and evaluating
defense methods against RNN adversarial attacks at all, and particularly
not in the cyber security domain, in which adversaries actually exist-
malware writers, who want their malware to evade the detection of
next generation, machine learning based malware classifiers. Our methods
reduces the number of adversarial examples evading the classifier
by more than 85\%.

\section{Background and Related Work}

\subsection{\label{subsec:RNN-Adversarial-Examples}RNN Adversarial Examples}

The search for adversarial examples is formalized as a minimization
problem \cite{DBLP:journals/corr/SzegedyZSBEGF13,Biggio2013}:

\begin{equation}
\arg_{\boldsymbol{r}}\min f(\boldsymbol{x}+\boldsymbol{r})\neq f(\boldsymbol{x})\:s.t.\:\boldsymbol{x}+\boldsymbol{r}\in\boldsymbol{D}
\end{equation}
The input \textbf{$\boldsymbol{x}$}, correctly classified by the
classifier $f$, is perturbed with \textbf{$\boldsymbol{r}$} such
that the resulting adversarial example \textbf{$\boldsymbol{x}+\boldsymbol{r}$}
remains in the input domain \textbf{$\boldsymbol{D}$} but is assigned
a different label than \textbf{$\boldsymbol{x}$}. To solve Equation
1, we need to transform the constraint $f(\boldsymbol{x}+\boldsymbol{r})\neq f(\boldsymbol{x})$
into an optimizable formulation. Then we can easily use the Lagrange
multiplier to solve it. To do this, we define a loss function $Loss()$
to quantify this constraint. This loss function can be the same as
the training loss, or it can be chosen differently, e.g., hinge loss
or cross-entropy loss.

Most sequence based adversarial attacks take place in the NLP domain.
Papernot et al. \cite{Papernot2016a} presented a white-box adversarial
example attack against RNNs, demonstrated against LSTM architecture,
for sentiment classification of a movie review dataset, where the
input is the review, and the output is whether the review was positive
or negative. The adversary iterates over the words \textbf{$\boldsymbol{x}[i]$}
in the review and modifies it as follows:

\begin{equation}
\boldsymbol{x}[i]=\arg\min_{\boldsymbol{z}}||sign(\boldsymbol{x}[i]-\boldsymbol{z})-sign(J_{f}(\boldsymbol{x})[i,f(\boldsymbol{x})])||
\end{equation}

\[
s.t.\:\boldsymbol{z}\in\boldsymbol{D}
\]

where $f(\boldsymbol{x})$ is the original model label for \textbf{$\boldsymbol{x}$},
and $J_{f}(\boldsymbol{x})[i,j]=\frac{\partial f_{j}}{\partial x_{i}}(\boldsymbol{x})$.
$sign(J_{f}(\boldsymbol{x})[i,f(\boldsymbol{x})])$ provides the direction
one has to perturb each of the word embedding components in order
to reduce the probability assigned to the current class and thus change
the class assigned to the sentence. The idea is that changing one
(or a few) word can change the meaning of the sentence. For instance,
changing: ``This is the \emph{best} movie have I ever seen'' to:
``This is the \emph{worst} movie I have ever seen.'' This approach
of modifying a word by the gradient is commonly used in many attacks
in order to achieve maximum classification impact with a minimal amount
of changes. Gao et al. \cite{JiDeepWordBug18} attacked sentiment
classification models in a black-box setting by either inserting,
deleting, or swapping characters to generate misspelled words mapped
into the 'unknown' word in the NLP dictionary, using various scoring
functions to find the most important words to modify. Other than attacking
text classifiers, Jian and Liang \cite{DBLP:conf/emnlp/JiaL17} aimed
to fool reading comprehension systems by adding misleading sentences.
Zhao et al. \cite{zhengli2018iclr} used a generative adversarial
network (GAN) to craft natural adversarial examples. Seq2seq models
are attacked in \cite{DBLP:conf/acl/EbrahimiRLD18,DBLP:journals/corr/abs-1803-01128},
which use a word-level attack method (the latter focuses on adding
specific ``malicious'' keywords to the adversarial sentence). Alzantot
et al. \cite{DBLP:conf/emnlp/AlzantotSEHSC18} presented an attack
algorithm that exploits population based gradient-free optimization
via genetic algorithms.

Attacks in the cyber security domain, mainly for malware classifiers
based on API calls, have also been presented. %
\begin{comment}
This domain presents some unique challenges: 1) The sequence length
can be extremely long (millions of API calls). 2) There is no such
thing as ``unknown API'' - usually the adversary is limited to the
valid dictionary (of possible APIs), making some effective NLP domain
attacks (e.g., \cite{JiDeepWordBug18}) infeasible. 3) The adversary
cannot perform any change he desires, because he wants to preserve
the (malicious) functionality of the modified malware. Despite those
challenges, TODO: Remove if we focus on cyber only
\end{comment}
{} Hu and Tan \cite{DBLP:journals/corr/HuT17a} presented a generative
RNN based approach, in which invalid APIs are generated and inserted
into the original API sequences. Recently, Rosenberg et al. \cite{DBLP:conf/raid/RosenbergSRE18}
presented a black-box variant of the attack in \cite{Papernot2016a},
by creating a substitute model and attacking it using a similar method,
and extended it to hybrid classifiers combining static and dynamic
features and architectures. Rosenberg et al. \cite{DBLP:journals/corr/abs-1804-08778}
further presented both black-box and white-box query-efficient attacks
based on perturbations generated using a GAN that was trained on benign
samples.

\subsection{\label{subsec:Defense-Mechanisms-against}Defense Mechanisms Against
Non-Sequence Based Adversarial Attacks}

Several methods have been suggested to detect whether a sample is
an adversarial example.

Some papers focused on statistical properties of adversarial examples.
Grosse et al.\cite{Grosse2017a} leveraged the fact that adversarial
samples usually have a different distribution than normal samples.
The statistical differences between them can be detected using a high-dimensional
statistical test of maximum mean discrepancy or by adding another
class of adversarial examples to the classifier. In contrast to our
work, their research deals with non-sequential input only. Metzen
et al. \cite{Metzen17} took a similar approach and augment deep neural
networks with a small \textquotedblleft detector\textquotedblright{}
subnetwork which is trained on the binary classification task of distinguishing
genuine data from data containing adversarial perturbations. Feinman
et al. \cite{Feinman17} detected adversarial examples using two new
features: kernel density estimates in the subspace of the last hidden
layer of the original classifier and Bayesian neural network uncertainty
estimates. Meng et al. \cite{DBLP:conf/ccs/MengC17} used a reformer
network (which is an auto-encoder or a collection of auto-encoders)
is trained to differentiate between normal and adversarial examples
by approximating the manifold of normal examples. When using a collection
of autoencoders, one reformer network is chosen at random at test
time, thus strengthening the defense.

In \cite{Carlini2017}, it was shown that most techniques like these
cannot handle a well-designed adversarial attack in the image recognition
domain.%
\begin{comment}
Moreover, in the cyber security domain, the question of what to do
after detecting an adversarial example remains open: Would such a
sequence be blocked? Be classified as malicious? This might increase
the FP rate, due to misclassifying samples as adversarial examples
and then treating them as malicious. Only reporting such cases is
also not an option in the cyber security domain, since this will allow
malicious files wrapped with the our framework (or an equivalent one)
to evade being blocked. Therefore, we would focus on the other sub-category
mentioned below.
\end{comment}

Xu et al. \cite{DBLP:conf/ndss/Xu0Q18} used a different approach,
\emph{feature squeezing}, to detect adversarial examples. This is
done by reducing the search space available to an adversary by coalescing
samples that correspond to many different feature vectors in the original
space into a single sample; this is accomplished by applying various
image-specific dimensionality reduction transformations to the input
features. If the original and squeezed inputs produce substantially
different outputs from the model, the input is likely to be adversarial.
This method applied the following image-specific dimensionality reduction
transformations to the input features: 1) Changing the image color
depth (e.g., from 24 bit to 8 bit). 2) Spatial smoothing (blur). However,
applying those transformations to discrete sequence input (e.g., API
call trace input for malware classification or words for sentiment
analysis) is not possible, because those transformation only fit images.

Instead of actively trying to detect adversarial examples, another
approach is to passively try to make the classifier more robust against
such attacks. Such methods avoid the false positives that might occur
in the abovementioned techniques. Using an ensemble of DNNs as a classifier
resistant to adversarial attacks on images was shown in \cite{Strauss17}.
In contrast to our work, this study only deals with feedforward networks
(mostly CNNs) in the computer vision domain. Stokes et al. \cite{DBLP:journals/corr/abs-1712-05919}
evaluate three defense methods: weight decay, ensemble of classifiers,
and distillation for a dynamic analysis malware classifier based on
a non-sequence based deep neural network. 

Some papers have also used GAN based approaches. Lee et al. \cite{DBLP:journals/corr/LeeHL17}
alternately trained both classifier and generator networks. The generator
network generates an adversarial perturbation that can easily fool
the classifier network by using a gradient of each image. Simultaneously,
the classifier network is trained to classify correctly both original
and adversarial images generated by the generator. These procedures
help the classifier network become more robust to adversarial perturbations.
Samangouei et al.\cite{DBLP:journals/corr/abs-1805-06605} trained
a GAN to model the distribution of unperturbed images. At inference
time, the closest output (which does not contain the adversarial changes)
to the input image is found. The generated image is then fed to the
classifier, and its prediction is used as the prediction of the original
input. In contrast to our work, this paper only deals with feedforward
networks (mostly CNNs) in the computer vision domain.

Adversarial training was suggested in \cite{DBLP:journals/corr/SzegedyZSBEGF13},
which demonstrated the injection of correctly labeled adversarial
samples in the training set as a means of making the model robust.
Tramer et al. \cite{DBLP:journals/corr/TramerKPBM17} introduced ensemble
adversarial training, a technique that augments training data with
perturbations transferred from other models.

To the best of our knowledge, there is currently no published and
evaluated method to make a \emph{sequence based RNN model} resistant
to adversarial sequences, beyond a brief mention of adversarial training
as a defense method \cite{DBLP:conf/emnlp/AlzantotSEHSC18,DBLP:journals/corr/abs-1812-05271}. 

Adversarial training has several limitations:
\begin{enumerate}
\item It provides a varying level of robustness, depending on the adversarial
examples used.
\item It requires a dataset of adversarial examples to train on. Thus, it
has limited generalization against novel adversarial attacks.
\item It requires retraining the model if the training set is large, potentially
incurring significant overhead.
\end{enumerate}
Our paper is the first to present and evaluate defense methods for
RNN classifiers, presenting five new defense methods and comparing
them to adversarial training.

\section{\label{sec:Methodology}Methodology}

In this paper we investigate six defense methods which are described
in the subsections that follow. An overview of the different defense
methods appear in Table \ref{tab:Defense-Method-Overview}.

\begin{table}[t]
\caption{\label{tab:Defense-Method-Overview}Defense Method Overview}
\centering{}%
\begin{tabular}{|>{\centering}p{0.15\textwidth}|>{\centering}p{0.12\textwidth}|>{\centering}p{0.08\textwidth}|}
\hline 
Defense Method & Attack-Specific/ Attack-Agnostic & Novelty\tabularnewline
\hline 
\hline 
Sequence Squeezing & Attack-Agnostic & Novel\tabularnewline
\hline 
\hline 
Defense Sequence-GAN & Attack-Agnostic & Inspired\tabularnewline
\hline 
\hline 
Nearest Neighbor & Attack-Agnostic & Inspired\tabularnewline
\hline 
\hline 
RNN Ensemble & Attack-Agnostic & Inspired\tabularnewline
\hline 
\hline 
Adversarial Signatures & Attack-Specific & Inspired\tabularnewline
\hline 
\hline 
Adversarial Training & Attack-Specific & Known\tabularnewline
\hline 
\end{tabular}
\end{table}

The evaluated defense methods are divided into three subgroups: Novel
RNN defense methods, known and previously evaluated RNN defense methods
and RNN defense methods inspired by existing CNN defense methods.
We implemented the latter subgroup ourselves as a baseline for our
novel attack. Each defense method is either \emph{attack-specific},
meaning it requires adversarial examples generated by the attack to
mitigate, or \emph{attack-agnostic}, that is, it works against all
types of attack methods, without the need to have a dataset of adversarial
examples generated by those attacks, making the latter a more preferable
choice.

Some of the suggested methods affect the classifier (such as the RNN
ensemble method), while others affect only the sequential input (e.g.,
the nearest neighbor method).

One might claim that some of the defense methods presented in this
paper are irrelevant, because there are inspired by CNN defense methods
that have been proven to be ineffective in cases where the attacker
is aware of the defense method being used and can devise a specialized
attack against this method. We call these attacks \emph{adaptive attacks}.
Such attacks have been published in \cite{Carlini2017} (against the
detection of adversarial examples using statistical irregularities),
in \cite{DBLP:conf/woot/HeWCCS17} (against feature squeezing) and
in \cite{DBLP:conf/icml/AthalyeC018,DBLP:conf/ccs/HashemiCK18,DBLP:journals/corr/abs-1712-09196}
(against Defense-GAN). However, implementing adaptive attacks in the
cyber security domain is more challenging than in the image recognition
domain, due to the following differences:
\begin{enumerate}
\item They are evaluated against variable length sequential discrete input
and against RNN classifiers, as opposed to continuous non sequential
fixed size input.
\item Modifying features in the cyber domain is only possible if the malicious
functionality remains intact following this modification.
\item An image used as input to an image classifier (usually a convolutional
neural network, CNN) is represented as a fixed size matrix of pixel
colors. If the actual image has different dimensions than the input
matrix, the picture will usually be resized, clipped, or padded to
fit the dimension limits. None of those transformation can be done
to a PE file while keeping its functionality intact.
\end{enumerate}
Furthermore, to the best of our knowledge, no CNN defense method published
so far is immune to breaking. Even if a defense method effectiveness
can be reduced by an adaptive attacker - it is still better than no
defense at all, since it defend against the simpler attacks and require
more effort from the attacker in order to implement an adaptive attack.

We didn't evaluate defense methods with inappropriate performance
(e.g., verifiable training, with 5\% test error for MNIST \cite{DBLP:conf/icml/WongK18})
and those which makes no sense or require significant modifications
to fit discrete sequential input based RNN classifiers (e.g., randomized
smoothing, which is certifiably robust under the $L_{2}$ norm \cite{DBLP:conf/icml/CohenRK19},
which makes less sense for sequential input).

\subsection{Threat Model}

We assume an adversary with full access to a trained target model,
with unlimited number of possible queries, so query efficient attack
(e.g., \cite{DBLP:journals/corr/abs-1804-08778}) is not an issue.
However, the adversary has no ability to influence that model. The
adversary tries to perturb malware to be misclassified by the model
using either black-box or white-box attack techniques, as specified
in Section \ref{subsec:Evaluated-Attacks}. The other sort of adversarial
attack (a benign sample being perturbed to be misclassified as malicious)
makes no sense in real-life, and this case is thus being ignored in
this paper. We also evaluate the cases where the adversary is aware
of the defense methods being used (adaptive white-box attacks in Section
\ref{subsec:Evaluated-Attacks}).

In this paper, we evaluate our defense methods against API call based
RNN malware classifiers, which, as previously mentioned, is a common
and concrete use case in the cyber security domain. However, all the
defense methods mentioned below are domain agnostic and can be used
in any domain with discrete sequence input. Evaluating those defense
in other domains would be part of our future work.

\subsection{Evaluated Defense Methods}

\subsubsection{Sequence Squeezing}

\emph{Sequence squeezing} is coalescing samples that correspond to
many different feature vectors in the original space into a single
vector that preserves the original meaning. If the original and squeezed
inputs produce substantially different outputs from the model, the
input is likely to be adversarial, and the features removed by the
squeezing might be the added adversarial perturbation. The squeezed
input is classified using the original classifier \emph{without retraining}
it, while reducing the search space available to an adversary by merging
semantic similar features into a single representative feature. 

For instance, a malware trying to communicate with a command and control
(CNC) server would prefer to use \emph{HttpSendRequestA()}. However,
this API is commonly used by malware and would be detected by malware
classifiers. Thus, an adversarial example would instead use \emph{HttpSendRequestW()},
not as commonly used by malware, in order to evade detection. Using
sequence squeezing, both \emph{HttpSendRequestA()} and \emph{HttpSendRequestW(),}
which have similar semantic meaning (here: functionality), would be
squeezed into a single feature group represented by \emph{HttpSendRequestA}(),
which better represents the group's semantic meaning due to its common
use. The classifier would see the \emph{HttpSendRequestA()} in the
input sequence instead of \emph{HttpSendRequestW(),} and this evasion
attack would be blocked.

Xu et al. used \emph{feature squeezing} \cite{DBLP:conf/ndss/Xu0Q18}
to detect adversarial examples for images. The essence of their method
is the application of the following image-specific dimensionality
reduction transformations to the input features: 1) Changing the image
color depth (e.g., from 24 bit to 8 bit). 2) Spatial smoothing (blur).
However, applying those transformations to discrete sequence input
(e.g., API call trace input for malware classification or words for
sentiment analysis) makes no sense, because those transformation only
fit images.

We therefore implement a different method that preserves the semantic
meaning of input sequences, while being generic enough to be applied
in diverse domains. While this paper focus on the cyber security domain,
our implementation fits other sequence based domains, including those
with larger vocabularies and more sophisticated adversarial attacks,
including semantic transformations, reordering, etc., such as NLP.
The experimental evaluation in the NLP domain would be part of our
future work

\begin{figure*}[tp]
\begin{centering}
\textsf{\includegraphics[scale=0.45]{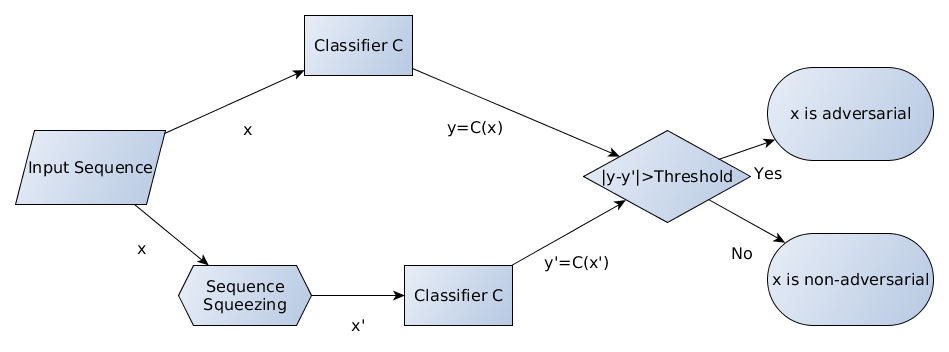}}
\par\end{centering}
\caption{\label{fig:Overview-of-the-2}Overview of the Sequence Squeezing Method}
\end{figure*}

Our method is illustrated in Figure \ref{fig:Overview-of-the-2} and
consists of the following stages:
\begin{enumerate}
\item Calculate word embedding, representing each API call/word in the vocabulary
in a semantic-preserving fashion, that is, words with similar meaning
have closer word embeddings.
\item Merge the words closest (=most similar) to a single center of mass
in order to reduce the dimensionality of the vocabulary, as well as
the adversarial space.
\item Replace all of the words that are part of the merged group with the
word closest to the center of mass of this group, maintaining lower
dimensionality and keeping the word embeddings used by the original
classifier. Thus, we can classify the squeezed input using the original
classifier without retraining it.
\item Apply the transformation (steps 1..3) on the classifier's input sequence.
If the classifier's confidence score for the squeezed sequence is
substantially different from the confidence score of the original
input (the threshold appears in Equation 3), the sample is adversarial. 
\end{enumerate}
The input sequence transformation is shown in Algorithm \ref{alg:Sequence-Squeezing-for}.

\begin{algorithm*}[tp]
\caption{\label{alg:Sequence-Squeezing-for}Sequence Squeezing for API Calls
and Words}
\begin{centering}
\textbf{Input}: $D$ (vocabulary), $X_{train}$ (training set of sequences
of tokens from $D$), $sizeD_{squeezed}$ (size of squeezed vocabulary)
\par\end{centering}
$embed=GloVe(D,X_{train})$ \# $embed\in\mathbb{R}^{|D|\times d}$

$newEmbed=embed$

\textbf{for} each word embedding \textbf{$\boldsymbol{\boldsymbol{w_{i}}}$}
in $embed$: 

\ \ \ \ $embedMergeHashTable[\boldsymbol{w_{i}}]=\left\{ \boldsymbol{w_{i}}\right\} $

\textbf{while} $(width(newEmbed)>sizeD_{squeezed})$:

\ \ \ \ \textbf{for} the two Euclidean closest word embeddings
in $newEmbed$, \textbf{$\boldsymbol{w_{i}},\boldsymbol{w_{j}}$}:

\ \ \ \ \ \ \ \ $\boldsymbol{w_{merged}}=\frac{\boldsymbol{w_{i}}*|embedMergeHashTable[\boldsymbol{w_{i}}]|+\boldsymbol{w_{j}}*|embedMergeHashTable[w_{j}]|}{|embedMergeHashTable[\boldsymbol{w_{i}}]|+|embedMergeHashTable[w_{j}]|}$ 

\ \ \ \ \ \ \ \ \# Calculate center of mass for the merged
embedding

\ \ \ \ \ \ \ \ $mergedGroup=embedMergeHashTable[\boldsymbol{w_{i}}]\cup embedMergeHashTable[w_{j}]$

\ \ \ \ \ \ \ \ \# The merged count is the aggregated count

\ \ \ \ \ \ \ \ remove \textbf{$\boldsymbol{w_{i}},\boldsymbol{w_{j}}$
}from\textbf{ }$newEmbed$ 

\ \ \ \ \ \ \ \ delete $embedMergeHashTable[\boldsymbol{w_{i}}]$

\ \ \ \ \ \ \ \ delete $embedMergeHashTable[w_{j}]$

\ \ \ \ \ \ \ \ \# Insert the merged embedding \textbf{$\boldsymbol{w_{merged}}$}
instead of \textbf{$\boldsymbol{w_{i}},\boldsymbol{w_{j}}$}

\ \ \ \ \ \ \ \ insert \textbf{$\boldsymbol{w_{merged}}$
}to\textbf{ }$newEmbed$

\ \ \ \ \ \ \ \ $embedMergeHashTable[\boldsymbol{w_{merged}}]=mergedGroup$

\ \ \ \ \ \ \ \ \# $width(newEmbed)$ was decremented by 1

\textbf{for} each word embedding \textbf{$\boldsymbol{w_{new}}$}
in $newEmbed$:

\ \ \ \ \textbf{if} $|embedMergeHashTable[\boldsymbol{\boldsymbol{w_{new}}}]|>1$

\ \ \ \ \# Meaning: $\boldsymbol{\boldsymbol{w_{new}}}$ is a
merged word embedding 

\ \ \ \ \ \ \ \ replace \textbf{$\boldsymbol{w_{new}}$} with
the Euclidean closest word embedding in \textbf{$embedMergeHashTable[\boldsymbol{\boldsymbol{w_{new}}}]$}

\textbf{return} $newEmbed$
\end{algorithm*}

We used GloVe \cite{pennington2014glove} word embedding, which, in
contrast to other methods (e.g., word2vec), has been shown to work
effectively with API call traces \cite{DBLP:conf/sigsoft/HenkelLLR18}%
\begin{comment}
TODO: 1. Explain more about GloVe (vs. word2vec? https://www.quora.com/How-is-GloVe-different-from-word2vec).
2. Move the reference to Section 4.2.1.
\end{comment}
, in order to generate $embed$, the word embedding matrix (of size:$|D|\times d$,
where $d$ is the embedding dimensionality) for each API call/word
(line 2). This word embedding is robust enough for domains with larger
vocabularies, to make our attack applicable to other domains such
as NLP .

We then perform agglomerative (bottom-up) hierarchical clustering
on the word embeddings, merging the closest word embedding (using
Euclidean distance, as in \cite{DBLP:conf/emnlp/AlzantotSEHSC18}
- line 7, as no significant improvement was observed when using cosine
distance. For word embedding, by definition, small Euclidean indeed
distance implies close semantic meaning). Each time we merge two embeddings,
we replace them with their center of mass, which becomes the embedding
of the merged group to which each of the merged embedding is mapped
(lines 10-17). This use of the center of mass preserves the overall
semantic of the group and allows us to keep merging the most semantically
similar words into groups (even to groups which were previously merged).

After the merging has been completed, we replace each merged group's
center of mass embedding with the closest (based on the Euclidean
distance) original word merged into it, so we can use the original
classifier trained on the original embeddings (line 16). The rationale
for this is that we want to choose the API or word with the closest
semantic meaning to the merged group members (represented by the merged
group's center of mass embedding), in order to maintain the performance
of the original classifier. 

To detect adversarial examples, we run the original classifier twice:
once for the original input and a second time for the sequence squeezed
input. If the difference of the confidence scores of the results is
larger than $Threshold_{adv}$, we say that the original input is
adversarial. We chose $Threshold_{adv}$ to be the maximum difference
between the original input and the squeezed input of all of the samples
in the training set. Thus, this is the minimal threshold that would
not affect the training set classification (and thus the original
classifier training). Additional details are available in Section
\ref{subsec:Sequence-Squeezing}.

\begin{equation}
Threshold_{adv}=\max(|f(\boldsymbol{x})-f(\boldsymbol{x_{squeezed}})|)
\end{equation}

\[
s.t.\:\boldsymbol{x}\in X_{train},
\]

\[
\boldsymbol{x_{squeezed}}=Algorithm1(D,X_{train},sizeD_{squeezed})[\boldsymbol{x}]
\]

The defense method training overhead is low and only involves iterating
the training set to calculate the squeezing transformation. No classifier
retraining with the squeezed vectors takes place. The original classifier
is still being used.

The inference overhead is also low: classifying each sample twice
(once with the original input sequence and a second time with the
squeezed sequence) using the original classifier.

\subsubsection{\label{subsec:Defense-SeqGAN}Defense Sequence-GAN}

One way to filter out the perturbations added by adversarial attack
is to train a GAN to model the distribution of unperturbed input.
At inference time, the GAN output (which does not contain the out-of-distribution
perturbation) that is closest to the target classifier's input is
found. This input is then fed to the classifier instead, and its prediction
is used as the prediction of the original input.

Samangouei et al. \cite{DBLP:journals/corr/abs-1805-06605} presented
Defense-GAN, in which a GAN was trained to model the distribution
of unperturbed images. However, Defense-GAN is defined for real-valued
data only, while API calls of a malware classifier are discrete symbols.
Small perturbations required by such GANs are not applicable to discrete
API calls.

For instance, you can't change \emph{WriteFile()} to \emph{WriteFile()}+0.001
in order to estimate the gradient needed to perturb the adversarial
example in the right direction; you need to modify it such that it
is an entirely different API.

The discrete outputs from the generative model make it difficult to
pass the gradient update from the discriminative model to the generative
model. We therefore used sequence GANs, i.e., GAN architectures designed
to produce sequences, in order to adapt this method for input sequences.

Several sequence GAN types were evaluated. For each sequence GAN type,
we trained a sequence GAN per class. In this study, this means that
a ``benign sequence GAN'' is used to produce API call sequences
drawn from the benign distribution used to train the GAN (the GAN
is trained using benign samples from the dataset; see Section \ref{subsec:Defense-Sequence-GAN}).
A ``malicious sequence GAN'' is used to produce malicious API call
sequences (the GAN is trained using malicious samples from the dataset;
see Section \ref{subsec:Defense-Sequence-GAN}).

\begin{figure*}[tp]
\begin{centering}
\textsf{\includegraphics[scale=0.45]{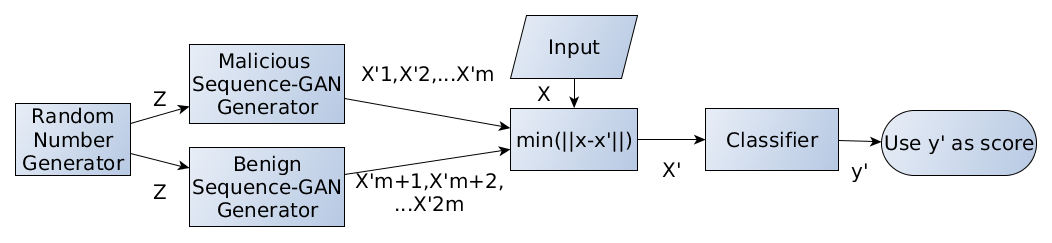}}
\par\end{centering}
\caption{\label{fig:Overview-of-the-1}Overview of the defense sequence GAN
Method}
\end{figure*}

For any input sequence to be classified, $m$ benign API call sequences
are generated by the ``benign GAN,'' and $m$ malicious API call
sequences are generated by the ``malicious GAN.'' We calculate the
Euclidean distance (no significant improvement was observed when using
the cosine distance) between the input and each of the $2m$ generated
sequences, choosing the sequence nearest to the original input sequence.
We then return the classifier's prediction for the nearest sequence
to the input sequence. The defense sequence GAN method is illustrated
in Figure \ref{fig:Overview-of-the-1}.

We have evaluated the following sequence GAN types: 

\paragraph{SeqGAN}

In SeqGAN \cite{DBLP:conf/aaai/YuZWY17} implementation, a discriminative
model is trained to minimize the binary classification loss between
real benign API call sequences and generated ones. In addition to
the pretraining procedure that uses the MLE (maximum likelihood estimation)
metric, the generator is modeled as a stochastic policy in reinforcement
learning (RL), bypassing the generator differentiation problem by
directly performing a gradient policy update. Given the API sequence
$s_{t}=\left[x_{0},x_{1},..x_{t-1}\right]$ and the next API to be
sampled from the model $x_{t}\sim\left(x|s_{t}\right)$, the RL algorithm,
REINFORCE, optimizes the GAN objective:

\begin{equation}
\underset{\phi}{\min}-\underset{Y\sim p_{data}}{E}\left[logD_{\phi}(Y)\right]-\underset{Y\sim G_{\Theta}}{E}\left[log(1-D_{\phi}(Y))\right]
\end{equation}
The RL reward signal comes from the GAN discriminator, judged on a
complete sequence, and is passed back to the intermediate state-action
steps using Monte Carlo search, in order to compute the Q-value for
generating each token. This approach is usedfor variance reduction.

\paragraph{TextGAN}

Zhang et al. \cite{DBLP:conf/icml/ZhangGFCHSC17} proposed a method
that optimizes the maximum mean discrepancy (MMD) loss, which is the
reconstructed feature distance, by adding a reconstruction term in
the objective.

\paragraph{GSGAN}

The Gumbel-Softmax trick is a reparametrization trick used to replace
the multinomial stochastic sampling in text generation \cite{DBLP:journals/corr/KusnerH16}.
GSGAN uses $argmax[softmax(h+g)]\backsim softmax(h)$, where $g$
is a Gumbel distribution with zero mean and unit variance. Since this
process is differentiable, backpropagation can be directly applied
to optimize the GAN objective.

\paragraph{MaliGAN}

The basic structure of MaliGAN \cite{DBLP:journals/corr/CheLZHLSB17}
follows that of SeqGAN. To stabilize the training and alleviate the
gradient saturating problem, MaliGAN rescales the reward in a batch.

In this case, the defense method training overhead is high: training
two GANs with stable performance is challenging, as discussed in \cite{Arjovsky17}.

The inference overhead is also high: generating $2m$ sequences using
the two GANs and finding the minimal distance from the input sequence.
This defense method's overhead may not be practical for cyber security
applications, where adversaries need to be identified in real time.

\subsubsection{\label{subsec:Nearest-Neighbor-Defense}Nearest Neighbor}

In this method, instead of returning the classifier score for the
input sequence, we return the score of the training set's sample nearest
to it, using the Euclidean distance (again, no significant improvement
was observed using the cosine distance).

This method leverages the fact that adversarial examples try to add
minimal perturbation to the input sample (Equation 1), so most parts
of the input sequence remain identical to the source (in this case,
malicious) class, and the distance from it would be minimal, thus
classifying the input sequence correctly despite the adversarial perturbation
(a discussion on the importance of minimal perturbation in the cyber
domain appears in Section \ref{subsec:Evaluated-Attacks}).

Note that this method is similar to the defense sequence-GAN method
(presented in Section \ref{subsec:Defense-SeqGAN}), however, instead
of using the GAN output closest to the input sequence, we use the
training set sample closest to the input sequence.

This defense method has no training overhead (the classifier training
does not change).

However, the inference overhead is high: finding the minimal distance
of the input sequence from all of the training set vectors. This is
especially true for large training sets, as used by most real world
commercial models.

This overhead can be reduced, e.g., by clustering all of the training
set vectors, and then calculating the distance to each sample in the
closest clusters. However, the overhead problem still exist for large
training sets. Another approach is using the input sequence distance
only from the centroids. However, this would negatively affect the
method's detection rate due to the lower granularity.

\subsubsection{\label{subsec:RNN-Ensemble}RNN Ensemble}

An ensemble of models represents a detour from the basic premise of
deep neural networks (including recurrent neural networks): training
a single classifier on all of the data to obtain the best performance,
while overfitting is handled using different mechanisms, such as dropout.

However, an ensemble of models can be used to mitigate adversarial
examples. An adversarial example is crafted to bypass a classifier
looking at all of the input. Thus, an ensemble of models, each focusing
on a subset of the input features, is more robust, since the models
trained on the input subsets would not be affected by perturbations
performed on other input subsets. Since adversarial examples are usually
constructed with a minimum amount of localized perturbations in the
input sequence (Equation 1; see discussion in Section \ref{subsec:Evaluated-Attacks}),
they would affect only a small number of models, but would be detected
by, e.g., ensemble majority voting.

The use of an ensemble of models as a defense method against adversarial
examples for images was suggested in \cite{Strauss17}. However, the
authors only presented the first two types of models mentioned in
our paper (regular and bagging) and only a single decision method
(hard voting), while we leverage ensemble models that provide better
accuracy for sequence based models, e.g., subsequence models.

We evaluate four types of ensemble models:
\begin{enumerate}
\item Regular models - Each model is trained on the entire training set
and all of the input sequences. The difference between the models
in the ensemble is due to the training method: each model would have
different initial (random) weights and the training optimizer that
can converge to a different function in each model, due to the optimizer's
``stochasticness'' (e.g., a stochastic gradient descent optimizer
picking different sample mini-batches and therefore converges to a
different loss function's local minimum of the neural network). This
means that each neural network learns a slightly different model.
\item Bagging models - Bagging \cite{Breiman:1996:BP:231986.231989} is
used on the training set. In those models, the training set consists
of drawing $|X_{train}|$ samples with replacement from the training
dataset of $|X_{train}|$ samples. On average, each model is trained
on 63.02\% of the training set, where many samples appear more than
once (are oversampled) in the new training set. \\
This means that each model is trained on a random subset of the training
set samples. Thus, each model learns a slightly different data distribution,
(hopefully) making it more robust to adversarial examples generated
to fit the data distribution of the entire dataset. \\
While our models were trained using dropout (see Appendix B), bagging
and dropout are not equivalent: Dropout does not filter entire samples
of the training set (only specific neurons from the neural network)
or oversample them, as bagging does. Dropout is also applied randomly
per epoch and per sample, while a bagging model's training set is
deterministic.
\item Adversarial models - We start from regular models ensemble mentioned
above. For each model in the ensemble, adversarial examples are generated
and the model is replaced with a model trained on the original model
training set and the generated adversarial examples. Thus, these are
actually regular models trained using adversarial training (see Section
\ref{subsec:Adversarial-Training}).
\item Subsequence models - Since the classifier's input is sequential, we
can train each model on a subset of the input sequence, starting from
a different offset. That is, if our model is being trained over sequences
of 200 API calls, we can split the model into 10 submodels: one on
API calls 1..100, the second on API calls 11..110, and the tenth on
API calls 101..200. \\
Note that the starting offsets can also be randomized per submodel,
instead of fixed as was done in our research. \\
The idea is that the models which classify an API trace of an adversarial
example in a subsequence without a perturbed part (i.e., a purely
malicious trace) would be classified correctly, while the perturbed
parts would be divided and analyzed separately, making it easier to
detect that the trace is not benign.
\end{enumerate}
Additional details are available in Section \ref{subsec:RNN-Ensemble-1}.

The output of the ensemble was calculated using one of two possible
methods:
\begin{enumerate}
\item Hard voting - Every model predicts its own classification, and the
final classification is selected by majority voting between the models
in the ensemble.
\item Soft voting - Every model calculates its own confidence score. The
average confidence score of all of the models in the ensemble is used
to determine the classification. Soft voting gives ``confident''
models in the ensemble more power than hard voting.
\end{enumerate}
This defense method does not require knowledge about the adversarial
examples during its setup in order to mitigate them, making it \emph{attack-agnostic},
with the exception of adversarial models, which are \emph{attack-specific},
based on the definition provided earlier in Section \ref{sec:Methodology}.

This defense method training overhead is high: training the number
of models in the ensemble instead of a single model.

The inference overhead is also high: running an inference for each
model in the ensemble instead of once.

\subsubsection{\label{subsec:Statistical-Sequence-Irregularit}Adversarial Signatures
(a.k.a. Statistical Sequence Irregularity Detection)}

Adversarial examples are frequently out-of-distribution samples. Since
the target classifier was not trained on samples from this distribution,
generalization to adversarial examples is difficult. However, this
different distribution can also be used to differentiate between adversarial
and non-adversarial samples. Our method searches for subsequences
of API calls that exist only (or mainly) in adversarial examples,
and not in regular samples, in order to detect if the sample is adversarial.
We call those subsequences \emph{adversarial signatures}.

Grosse et al. \cite{Grosse2017a} leverage the fact that adversarial
samples have a different distribution than normal samples for \emph{non-sequential}
input. The statistical differences between them can be detected using
a high-dimensional statistical test of maximum mean discrepancy. In
contrast, our method handles sequential input and leverages the conditional
probabilities between the sequence elements (API calls or words) instead
of the maximum mean discrepancy.

In order to do that, we start from the observation that in an API
call trace, as well as in natural language sentences, there is a strong
dependence between the sequence elements. The reason for this is that
an API call (or a word in NLP) is rarely independent, and in order
to produce usable business logic, a sequence of API calls (each relying
on the previous API calls' output and functionality) must be implemented.
For instance, the API call \emph{closesocket()} would appear only
after the API call \emph{socket()}. %
\begin{comment}
Using the statistical correlation between API calls is nothing new,
and was used to detect anomalous system call traces which are indicative
of a malicious process more than 20 years ago \cite{Forresta}.
\end{comment}
{} The same is true for sentences: an adverb would follow a verb, etc.

For most state-of-the-art adversarial examples, only a small fraction
of API calls is added to the original malicious trace (see discussion
about minimal perturbation in the cyber security domain in Section
\ref{subsec:Evaluated-Attacks}), and the malicious context (the original
surrounding API calls of the original business logic) remains. Thus,
we evaluated the probability of a certain API call subsequences to
appear, generating ``signatures'' of API call subsequences that
are more likely to appear in adversarial sequences, since they contain
API calls (the adversarial-added API calls) unrelated to their context.

We decided to analyze the statistical irregularities in n-grams of
consecutive API calls. The trade-off when choosing n is to have a
long enough n-gram to capture the irregularity in the proper context
(surrounding API calls), while remaining short enough to allow generalization
to other adversarial examples.

For each unique API call (the features used in \cite{DBLP:conf/raid/RosenbergSRE18,DBLP:journals/corr/HuT17a})
n-gram, we calculate the \emph{adversarial n-gram probability} of
the n-gram of monitored API calls $(w_{1},w_{2}..,w_{n})|\{w_{1},w_{2}..,w_{n}\}\subseteq D$,
where $D$ is the vocabulary of available features. Here those features
are all of the API calls recorded by the classifier.

\begin{equation}
p_{adv}(w_{1},w_{2}..,w_{n})=
\end{equation}

\[
\frac{|\left\{ x|x\in X_{adv}\wedge((w_{1}\perp w_{2}\perp..w_{n})\subseteq x)\right\} |}{|\left\{ x|x\in\left(X_{adv}\cup X_{train,target}\right)\wedge((w_{1}\perp w_{2}\perp..w_{n})\subseteq x)\right\} |}
\]

$\perp$ is the concatenation operation. The adversarial n-gram probability
is the ratio of occurrences of the n-gram in the adversarial example
dataset available to the defender $X_{adv}$, as part of the occurrences
in both the adversarial examples and target (i.e., benign) class samples
in the training set, $X_{train,target}$.

Note that the equation is valid regardless of $|X_{adv}|$, and there
is no assumption regarding the ratio between $|X_{adv}|$ and $|X_{train,target}|$. 

The reason we don't include malicious samples is that we want statistical
irregularities from the target class, which is the benign class in
this case. Also note that we only look at the appearance of the signatures
in the target class and not in other classes (i.e., we look only at
the benign class and not the malicious class). The reason for this
is that it makes sense that $X_{adv}$ would contain API n-grams available
in the source class (the malicious class in this case), because in
practice, it is a source class sample.

We say that the n-gram of monitored API calls $(w_{1},w_{2}..,w_{n})$
is an \emph{adversarial signature} if the adversarial n-gram probability
of this n-gram is larger than a threshold $ThresholdP_{adv}$ that
is determined by the trade-off between the adversarial example detection
rate and the number of target class samples falsely detected as adversarial;
the higher the threshold, the lower both would be.

We classify a sample as an adversarial example if it contains more
than $ThresholdSigs_{adv}$ adversarial signatures. The more irregular
n-grams detected, the more likely the sequence is to be adversarial.
Additional details are provided in Section \ref{subsec:Adversarial-Signatures}.

This defense method requires a dataset of adversarial examples, $X_{adv}$,
during its setup, in order to make it robust against such examples,
making it \emph{attack-specific}, based on the definition provided
earlier in Section \ref{sec:Methodology}.

Note that while finding ``non-adversarial signatures'' using this
method is possible, it is more problematic, especially when $|X_{train,target}|$
is very large. Other methods presented in this paper, such as defense
sequence GAN (see Section \ref{subsec:Defense-SeqGAN}), implement
this approach more efficiently.

This defense method training overhead is high: counting all subsequences
of a certain size in the training set.

The inference overhead, however, is low: searching for the adversarial
signatures in the input sequence.

\subsubsection{\label{subsec:Adversarial-Training}Adversarial Training}

Adversarial training is the method of adding adversarial examples,
with their non perturbed label (source class label), to the training
set of the classifier. The rationale for this is that since adversarial
examples are usually out-of-distribution samples, inserting them into
the training set would cause the classifier to learn the entire training
set distribution, including the adversarial examples.

Additional details are available in Section \ref{subsec:Adversarial-Training-1}.

Unlike all other methods mentioned in this paper, this method has
already been tried for sequence based input in the NLP domain (\cite{DBLP:conf/emnlp/AlzantotSEHSC18,DBLP:journals/corr/abs-1812-05271}),
with mixed results about the robustness it provides against adversarial
attacks. Additional issues regarding this method are described in
Section \ref{subsec:Defense-Mechanisms-against}. We evaluate this
method in order to determine whether the cyber security domain, with
a much smaller dictionary (less than 400 API call types monitored
in \cite{DBLP:conf/raid/RosenbergSRE18} compared to the millions
of possible words in NLP domains), would yield different results.
We also want to compare it to the defense methods presented in this
paper, using the same training set, classifier, etc.

This defense method training overhead is high (generating many adversarial
examples, following by training a classifier on a training set containing
them).

There is no inference overhead: the inference is simply performed
using the newly trained classifier.

\section{\label{sec:Experimental-Evaluation}Experimental Evaluation}

\subsection{\label{subsec:Dataset-and-Target}Dataset and Target Malware Classifiers}

We use the same dataset used in \cite{DBLP:conf/raid/RosenbergSRE18},
because of its size: it contains 500,000 files (250,000 benign samples
and 250,000 malware samples), faithfully representing the malware
families in the wild and allowing us a proper setting for attack and
defense method comparison. Details are provided in Appendix A. 

Each sample was run in Cuckoo Sandbox, a malware analysis system,
for two minutes per sample. Tracing only the first few seconds of
a program execution might not allow the detection of certain malware
types, like ``logic bombs'' that commence their malicious behavior
only after the program has been running for some time. However, this
can be mitigated both by classifying the suspension mechanism as malicious,
if accurate, or by tracing the code operation throughout the program
execution lifetime, not just when the program starts. The API call
sequences generated by the inspected code during its execution were
extracted from the JSON file generated by Cuckoo Sandbox. The extracted
API call sequences were used as the malware classifier's features. 

The samples were run on Windows 8.1 OS, since most malware targets
the Windows OS. Anti-sandbox malware was filtered to prevent dataset
contamination (see Appendix A). After filtering, the final training
set size is 360,000 samples, 36,000 of which serve as the validation
set. The test set size is 36,000 samples. All sets were balanced between
malicious and benign samples, with the same malware families composition
as in the training set.

There are no commercial trial version or open-source API call based
deep learning intrusion detection systems available (such commercial
products target enterprises and involve supervised server installation).
Dynamic models are also not available in free online malware scanning
services like \href{https://www.virustotal.com}{VirusTotal}. Therefore,
we used RNN based malware classifiers, trained on the API call traces
generated by the abovementioned dataset. 

The API call sequences were split into windows of $m$ API calls each,
and each window was classified in turn. Thus, the input of all of
the classifiers was a vector of $m=140$ (larger window sizes didn't
improve the classifier's accuracy) API call types, each with 314 possible
values (those monitored by Cuckoo Sandbox). The classifiers used to
evaluate the attacks are similar to those used in \cite{DBLP:conf/raid/RosenbergSRE18}.
Their implementation and hyperparameters (loss function, dropout,
activation functions, etc.), and the performance of the target classifiers
are described in Appendix B. 

\subsection{\label{subsec:Evaluated-Attacks}Evaluated Attacks}

The attacks used to assess the defense methods' robustness are described
in Appendix C. These are the state-of-the-art out of the few published
RNN attacks in the cyber security domain. The attacks add API calls
to the API trace (not removing or modifying API calls, in order to
avoid damaging the modified code's functionality) based on either
their gradients, maximizing the effect of each added API call, or
randomly. The maximum number of allowed adversarial API calls is 93
in each sliding window of $n=140$ API calls (i.e., 66.67\% which
is a very permissive boundary). 

Three attacks are used to evaluate the robustness of our defense methods:
\begin{enumerate}
\item A realistic gradient based black-box attack, in which the attacker
has no knowledge of the target classifier's architecture or weights
and has to build a substitute model, as done in \cite{DBLP:conf/raid/RosenbergSRE18}.
The holdout dataset size to build the substitute model was identical
(70 samples) for a fair comparison. The attack effectiveness for the
LSTM classifier (without any defense method) is 99.99\%.
\item A white-box gradient based attack, where the adversary uses the target
classifier instead of a substitute model. This attack is a stronger
variant of the attack used in \cite{DBLP:journals/corr/abs-1804-08778},
which has access only to the confidence score of the classifier. The
attack effectiveness for the LSTM classifier (without any defense
method) is 100.00\%
\item An adaptive attacker white-box attack, where the attacker is aware
not only of the classifier's architecture and hyperparameters, but
also of the defense method being used. Thus, the attacker operates
an adversarial attack specially crafted for the used defense method.
These attacks are described per defense method in Section \ref{subsec:Defense-Methods-Implementation}.
All of these attacks' effectiveness for the LSTM classifier is 100\%.
\item A random perturbation attack. The attack effectiveness for the LSTM
classifier (without any defense method) is 22.97\% (average of five
runs). 
\end{enumerate}
Adversarial attacks against images usually try to minimize the number
of modified pixels in order to evade human detection of the perturbation.
One might claim that such a definition of minimal perturbation is
irrelevant for API call traces: humans cannot inspect sequences of
thousands or millions of APIs, so an attacker can add an unlimited
amount of API calls, this improving the attack effectiveness against
the evaluated defense methods.

However, one should bear in mind that malware aims to perform its
malicious functionality as quickly as possible. For instance, ransomware
usually starts by encrypting the most critical files (e.g., the 'My
Documents' folder) first, so even if the user turns off the computer
and sends the hard drive to the IT department - damage has already
been done. The same is true for a keylogger that aims to send the
user passwords to the attacker as soon as possible, so they can be
used immediately, before the malware has been detected and neutralized.

Moreover, adding too many API calls would cause the program's profile
to become anomalous, making it easier to detect by anomaly detection
intrusion detection systems, e.g., those that measure CPU usage \cite{DBLP:conf/isi/MoskovitchPGSFPSE07},
or contain irregular API call subsequences (Section \ref{subsec:Statistical-Sequence-Irregularit}).

Finally, the robustness of the defense methods to the addition of
many API calls is being evaluated by the random perturbation attack.
The random attack adds 50\% API calls to the trace, as opposed to
0.0005\% API calls for the entire trace for the gradient based black-box
and white-box attacks \cite{DBLP:conf/raid/RosenbergSRE18}. or 10,000
times more API calls, buy has much lower evasion rate. Thus, samples
whose adversarial variants are detected cannot evade detection simply
by adding more perturbations.

\subsection{\label{subsec:Defense-Methods-Implementation}Defense Methods Implementation
Details}

Additional details about the implementation of the defense methods
are provided in the subsections that follow. Each subsection also
contains a description of the best adaptive white-box attack against
this method, that is, if the attacker knows this method is being used,
how can he/she bypass it most effectively.

\subsubsection{\label{subsec:Sequence-Squeezing}Sequence Squeezing}

We used Stanford's GloVe implementation \cite{GloveNLP} with embedding
dimensionality of $d=32$. The vocabulary $D$ used by our malware
classifiers contains all of the API calls monitored by Cuckoo Sandbox,
documented in the Cuckoo Sandbox repository \cite{CuckooHookedAPIs}. 

Running Algorithm \ref{alg:Sequence-Squeezing-for} on our training
set (see Section \ref{subsec:Dataset-and-Target}) of API call traces
resulted in interesting sequence squeezing. It seems that the squeezed
groups maintained the ``API contextual semantics'' as expected,
merging, for instance, the variants of the same API, e.g., \emph{GetUserNameA()}
and \emph{GetUserNameW()}. Other merged API calls are different API
calls with the same functionality, e.g., \emph{socket()} and \emph{WSASocketA()}.
\begin{comment}
TODO: Extend, either here or in Appendix D
\end{comment}

The sequence squeezing we used, with $sizeD_{squeezed}=\left\lceil \frac{|D|}{2}\right\rceil =157$,
is described in Appendix D. Using smaller feature spaces, e.g., $sizeD_{squeezed}=\left\lceil \frac{|D|}{3}\right\rceil $,
resulted in the inability to maintain the ``API contextual semantics,''
merging unrelated API calls and reducing the classifier's accuracy
by 7\%. However, using larder feature spaces, e.g., $sizeD_{squeezed}=\left\lceil \frac{2|D|}{3}\right\rceil $,did
not limit the adversarial space enough, resulting in 5\% adversarial
recall loss. For our training set, we used $Threshold_{adv}=0.18$. 

Other hyperparameters (grid-search selected) were less effective. 

\paragraph{Best Adaptive White-Box Attack}

To bypass sequence squeezing defense method, the attacker performs
the white-box attack (described in Section \ref{subsec:Evaluated-Attacks})
twice: once on the original input sequence (generating the adversarial
example $\boldsymbol{x}^{*}$) and once on the squeezed sequence (generating
the adversarial example $\boldsymbol{x_{squeezed}^{*}}$), both of
them using perturbation only from the squeezed vocabulary $D_{squeezed}$,
and not from $D$. This is done iteratively, until we bypass the threshold
being set in Equation 3: $Threshold_{adv}\geq(|f(\boldsymbol{x}^{*})-f(\boldsymbol{x_{squeezed}^{*}})|)$.
If this attack doesn't succeed after 10 iterations, the attack has
failed.

\subsubsection{\label{subsec:Defense-Sequence-GAN}Defense Sequence-GAN}

To implement the benign perturbation GAN, we tested several GAN types,
using Texygen \cite{DBLP:conf/sigir/ZhuLZGZWY18} with its default
parameters. We used MLE training as the pretraining process for all
of the GAN types except GSGAN, which requires no pretraining. In pretraining,
we first trained 80 epochs for a generator and then trained 80 epochs
for a discriminator. The adversarial training came next. In each adversarial
epoch, we updated the generator once and then updated the discriminator
for 15 mini-batch gradients. We generated a window of 140 API calls,
each with 314 possible API call types, in each iteration. 

As mentioned in Section \ref{subsec:Defense-SeqGAN}, we tested several
GAN implementations with discrete sequence output: SeqGAN \cite{DBLP:conf/aaai/YuZWY17},
TextGAN \cite{DBLP:conf/icml/ZhangGFCHSC17}, GSGAN \cite{DBLP:journals/corr/KusnerH16},
and MaliGAN \cite{DBLP:journals/corr/CheLZHLSB17}. We trained our
``benign GAN'' using a benign holdout set (3,000 sequences). Next,
we generated $m=200$ sequences with the ``benign GAN,'' using an
additional benign hold-out set (3,000 sequences) as a test set. We
used the same procedure to train our ``malicious GAN'' and generated
an additional $m=200$ sequences using it.

SeqGAN outperforms all other models by providing the average minimal
distance (both the Euclidean distance and cosine distance provided
similar results) between the 400 sequences generated and the test
set vectors, meaning that the sequences generated were the closest
to the requested distribution, thus, we used SeqGAN.

\paragraph{Best Adaptive White-Box Attack}

To bypass defense sequence GAN, the attacker performs the white-box
attack (described in Section \ref{subsec:Evaluated-Attacks}) and
then performs the same process done by the defender in Section \ref{subsec:Defense-SeqGAN}:
He/she generates $m=200$ sequences with the ``benign GAN,'' and
generates an additional $m=200$ sequences using the ``malicious
GAN''. The attacker iteratively repeats this process until the adversarial
sequence is closer to one of the benign GAN sequences which are classified
as benign. If this attack doesn't succeed after 10 iterations, the
attack has failed.

\subsubsection{\label{subsec:Nearest-Neighbor}Nearest Neighbor}

The cosine distance is more effective than the Euclidean distance
in many NLP tasks. However, the differences in performance due to
the use of cosine distance instead of Euclidean distance were marginal,
as mentioned in Section \ref{subsec:Nearest-Neighbor-Defense}. Therefore,
we used the Euclidean distance for nearest neighbor calculations.

\begin{comment}
TODOs: 1.. Add \cite{DBLP:conf/emnlp/AlzantotSEHSC18} attack (N:P
domain, non gradient based attack) 2. Ensemble (combinations) of 2
or more of the mentioned methods
\end{comment}

\paragraph{Best Adaptive White-Box Attack}

To bypass the nearest neighbor defense method, the attacker iteratively
performs the white-box attack (described in Section \ref{subsec:Evaluated-Attacks})
and then calculates the distance to the nearest neighbor in the training
set, until this neighbor's classification is benign. If this attack
doesn't succeed after 10 iterations, the attack has failed.

\subsubsection{\label{subsec:RNN-Ensemble-1}RNN Ensemble}

We used six variants of ensembles, each consisting of nine models:
\begin{itemize}
\item Regular ensemble - Each model was trained on the entire dataset.
\item Subsequence ensemble - The first model is trained on API calls at
offsets between 1..140 in the API call trace, the second model is
trained on API calls at offsets 11...150 in the API call trace, etc.
The ninth model is trained on API calls at offsets 91..230 in the
API call trace.
\item Bagging ensemble - Each model is trained on a random subset of the
dataset, as discussed in Section \ref{subsec:RNN-Ensemble}.
\item Bagging subsequence ensemble - This is a combination of the bagging
and subsequence ensembles. Each model is trained not only on a different
API call offset range (like the regular subsequence ensemble described
above) but also on a random subset of the training set, as in the
bagging ensemble.
\item Adversarial ensemble - Each model has 14,000 (out of 27,000) malicious
samples replaced with their adversarial example variants, similar
to adversarial training (see Section \ref{subsec:Adversarial-Training-1}).
\item Adversarial subsequence ensemble - This is a combination of the adversarial
and subsequence ensembles. The adversarial examples' API call traces
used for training also start at an offset.
\end{itemize}
The output of the ensemble presented in Table \ref{tab:Defense-Methods-Performance}
was made using soft voting (see Section \ref{subsec:RNN-Ensemble}).
However, the difference between soft and hard voting in all of our
tests was marginal.

\paragraph{Best Adaptive White-Box Attack}

To bypass RNN ensemble defense method, the attacker performs the white-box
attack (described in Section \ref{subsec:Evaluated-Attacks}) on each
classifier in the ensemble in turn, and aggregates the added adversarial
API calls from all the attacks into the same adversarial sequence.

\subsubsection{\label{subsec:Adversarial-Signatures}Adversarial Signatures}

We chose $n=5$, i.e., we used 5-grams of API calls for the ``adversarial
signatures.'' Shorter API call subsequences caused more ``adversarial
false positives,'' i.e., identifying regular samples as adversarial
(e.g., $n=4$ resulted in 0.5\% of the test set incorrectly classified
as adversarial), while longer API call sequences were too specific
to detect the majority of the adversarial examples (e.g., $n=6$,
while having optimal performance in other systems \cite{DBLP:conf/sp/TanM02},
resulted in a 15\% loss of adversarial recall). 

We used $ThresholdP_{adv}=1.0$, thus, we looked for API subsequences
which appear \emph{only} in adversarial examples. We also used $ThresholdSigs_{adv}=1$,
so that in order to classify a sample as an adversarial example, it
is enough that it contains a single \emph{adversarial signature}. 

Other hyperparameters (grid search selected) were less effective.

\paragraph{Best Adaptive White-Box Attack}

To bypass adversarial signatures, the attacker performs the white-box
attack (described in Section \ref{subsec:Evaluated-Attacks}), but
verifies for each added adversarial API call that its addition generates
no adversarial signatures. If it does, the attacker would choose a
different position in the sequence instead to add a different adversarial
API call, by re-calculating the gradients, as described in Appendix
C.

\subsubsection{\label{subsec:Adversarial-Training-1}Adversarial Training}

We ran the adversarial black-box and white-box attacks (see Appendix
C), as suggested in \cite{DBLP:conf/iclr/MadryMSTV18}, on the training
set. We generated 14,000 malicious adversarial examples (50\% generated
by the black-box attack and 50\% by the white-box attack), which replaced
14,000 malicious samples in the original training set. Other sizes
(grid search selected) resulted in reduced accuracy.

\paragraph{Best Adaptive White-Box Attack}

To bypass adversarial training, the attacker performs the white-box
attack (described in Section \ref{subsec:Evaluated-Attacks}) on the
classifier trained on the augmented training set that includes the
adversarial examples, either.

\subsection{\label{subsec:Defense-Methods-Performance}Defense Method Performance}

The different defense methods mentioned in Section \ref{sec:Methodology}
are measured using two factors:

The \emph{adversarial recall} is the fraction of adversarial sequences
generated by the attack which were either detected by the defense
method or classified as malicious by the classifier. Adversarial recall
provides a metric for the robustness of the classifier combined with
a defense method against a specific adversarial attack. 

The \emph{classifiers' accuracy}, which applies equal weight to both
false positives and false negatives (unlike precision or recall),
thereby providing an unbiased overall performance indicator. The accuracy
is evaluated against the regular test set only (that is, without adversarial
examples). Using this method, we verify the effect our defense methods
have on the classifiers' performance for non-adversarial samples.
Since samples classified as adversarial examples are automatically
classified as malicious (because there is no reason for a benign adversarial
example), every benign sample classified as an adversarial example
would damage the classifier's accuracy. Therefore, we also specify
the false positive rate of the classifier to better asses the effect
of those cases.

The performance of the attack versus the various defense methods and
the classifier's performance using those defense methods for the LSTM
classifier (see Appendix B; note that the other tested classifiers
mentioned in Appendix B behave the same) are presented in Table \ref{tab:Defense-Methods-Performance}.
All adaptive white-box attacks were tried against all defense methods,
but only the best one (with the lowest adversarial recall) is shown
in the table, due-to space limits.

The overhead column contains high-level observations regarding the
defense method's overhead (none, low, or high), both during training
(time and money) and during inference (run-time performance). The
analysis is described in detail in the subsections of Section \ref{sec:Methodology}.
As models get more complicated and larger in size, we would expect
defense methods with lower overhead to be more easily deployed in
real-world scenarios.

\begin{table*}[tbph]
\caption{\label{tab:Defense-Methods-Performance}Defense Method Performance}
\centering{}%
\begin{tabular}{|>{\centering}p{0.25\textwidth}|>{\centering}p{0.08\textwidth}|>{\centering}p{0.08\textwidth}|>{\centering}p{0.08\textwidth}|>{\centering}p{0.12\textwidth}|>{\centering}p{0.12\textwidth}|>{\centering}p{0.13\textwidth}|}
\hline 
Defense Method & Adversarial Recall {[}\%{]}, Best Adaptive White-Box Attack & Adversarial Recall {[}\%{]}, White-Box Attack & Adversarial Recall {[}\%{]}, Black-Box Attack & Adversarial Recall {[}\%{]}, Random Perturbation Attack (Average of
5 Runs) & Classifier Accuracy/False Positive Rate {[}\%{]}, Non-Perturbed Test
Set & Method Overhead (Training/

Inference)\tabularnewline
\hline 
Original Classifier (No Defense Methods) & - & - & - & - & 91.81/0.89 & None/None\tabularnewline
\hline 
\hline 
Sequence Squeezing & 38.76 & 86.96 & 48.57 & 99.14 & 90.81/0.99 & Low/Low\tabularnewline
\hline 
Defense Sequence-GAN & 13.18 & 50.49 & 57.38 & 97.35 & 64.55/3.85 & High/High\tabularnewline
\hline 
Nearest Neighbor & 11.68 & 88.78 & 56.12 & 40.56 & 87.85/1.32 & None/High\tabularnewline
\hline 
\hline 
RNN Ensemble (Regular) & 19.65 & 47.91 & 73.0 & 84.54 & 92.59/0.81 & High/High\tabularnewline
\hline 
\hline 
RNN Ensemble (Subsequence) & 22.72 & 86.27 & 56.79 & 85.55 & 92.27/0.84 & High/High\tabularnewline
\hline 
\hline 
RNN Ensemble (Bagging) & 18.23 & 47.90 & 73.01 & 87.29 & 92.65/0.80 & High/High\tabularnewline
\hline 
\hline 
RNN Ensemble (Bagging Subsequence) & 24.15 & 86.26 & 56.78 & 79.00 & 90.54/1.03 & High/High\tabularnewline
\hline 
\hline 
RNN Ensemble (Adversarial) & 21.17 & 47.92 & 73.02 & 99.57 & 92.97/0.76 & High/High\tabularnewline
\hline 
\hline 
RNN Ensemble (Adversarial Subsequence) & 27.19 & 86.27 & 56.79 & 99.34 & 92.21/0.85 & High/High\tabularnewline
\hline 
\hline 
Adversarial Signatures & 11.65 & 75.36 & 75.38 & 99.78 & 91.81/0.89 & High/Low\tabularnewline
\hline 
\hline 
Adversarial Training & 11.16 & 48.32 & 12.39 & 99.09 & 91.81/0.91 & High/None\tabularnewline
\hline 
\end{tabular}
\end{table*}

We see that certain defense methods, such as RNN subsequence ensembles
and adversarial signatures, provide good robustness against adversarial
attacks (75-85\% adversarial recall). Ensemble models also improve
the classifier performance on non-adversarial samples. However, our
novel defense method, sequence squeezing, provides a much lower overhead
than all the other methods, while still offering good robustness against
adversarial attacks, making it the most suited for real world scenarios
and especially when real time classification is required, like in
the cyber security domain. Sequence squeezing is also the best method
against an adaptive attacker, leaving the defender better off than
with any other evaluated defense method.

\subsection{Discussion}

Table \ref{tab:Defense-Methods-Performance} reveals several interesting
issues:

We see that the best defense method selection depends on the specific
scenario:
\begin{itemize}
\item When the training and inference overhead is not a concern, an RNN
subsequence ensemble provide good robustness against adversarial attacks
(especially white-box attacks, both adaptive and not) while improving
the classifier performance on non-adversarial samples. 
\item When an attack-specific defense is acceptable, adversarial signatures
provide good robustness against adversarial attacks (especially black-box
attacks), with lower overhead. 
\item Finally, for cases where a low overhead attack-agnostic defense method
is required (as required in most real world scenarios), the novel
sequence squeezing defense method provides good robustness against
adversarial attacks, with much lower overhead, at a price of a slight
degradation in the classifier performance against non-adversarial
samples (0.1\% addition to the false positive rate). This means that
sequence squeezing is the best choice for cyber security applications,
where on the one hand, adversaries need to be identified in real time,
but on the other hand, the adversarial attack method is unknown, so
attack-specific method like adversarial signatures is irrelevant.
\end{itemize}
Attack-specific defense methods (adversarial signatures, adversarial
training) present very poor adversarial recall against the best adaptive
white-box attack. Attack-agnostic defense methods which rely on the
assumption that the attacker would use minimal perturbation (defense
sequence GAN, nearest neighbor) provide slightly better performance,
but still cannot cope with attackers that use larger perturbations.
Our novel defense method, sequence squeezing, is the best one against
adaptive attacks. The reason is sequence squeezing reduces the dimensionality
of the possible adversarial perturbations, limiting the attacker's
possibilities to perturb the input and evade detection,

The evaluated defense methods usually provide greater robustness against
white-box attacks than against black-box attacks. This suggests that
transferable adversarial examples, the ones that bypass the substitute
model and are also effective against another classifier (the target
model), as done in the black-box attack, are different from other
adversarial examples (e.g., drawn from a different distribution),
causing defense methods to be less effective against them. For instance,
transferable adversarial examples might contain adversarial API calls
with similar semantic meaning to the API calls of the original benign
sample, making sequence squeezing less effective. Analyzing the differences
between black-box and white-box RNN adversarial examples and their
effect on our proposed defense methods will be addressed in future
work.

All of the evaluated defense methods provide adequate robustness against
a random perturbation attack. However, as shown earlier in Section
\ref{subsec:Defense-Methods-Performance}, this attack is less effective
than other attack types and is ineffective even when no defense method
is applied.

The accuracy of the RNN ensembles \textbf{is higher than the original
classifier} for all ensemble types. The RNN subsequence ensembles
also provide good robustness against white-box attacks, while the
non-subsequence ensembles provide good robustness against black-box
attacks. Bagging and adversarial models do not affect the performance
when adding them to the subsequence or regular models. \\
On the other hand, an ensemble requires the training of many models.
This affects both the training time and the prediction time: An ensemble
of nine models like in this study, requires nine times the training
and inference time of a single model. \\
One might claim that the improvement of RNN subsequence ensemble classifiers
on non-adversarial samples results from the fact that those classifiers
consider more API calls in each sliding window (230 API calls in all
nine models, instead of 140 API calls in a single model). However,
testing larger API call sliding windows for a single classifier (1000
API calls in \cite{DBLP:conf/raid/RosenbergSRE18}) did not result
in a similar improvement. This leads us to believe that the improved
performance actually derives from the ensemble itself and not from
the input size. This is supported by the improvement in non-subsequence
ensembles.

The sequence squeezing defense method provides good robustness against
white-box attacks but is less effective against black-box attacks.
This defense method causes a small decrease in the classifier performance
(about 1\%) and doubles the prediction time, because two models are
being run: the original and the sequence squeezed models.

The adversarial signatures defense method provides the best black-box
attack robustness, but its white-box attack performance is lower than
sequence squeezing.

The only published defense method to date, adversarial training, underperforms
in comparison to all other defense methods evaluated, since its adversarial
recall is lower.

For the defense sequence GAN method, the inability of the sequence
based GANs to capture the complex input sequence distribution (demonstrated,
for instance, by the fact that some of the API sequences generated
by the GAN contain only a single API call type, as opposed to the
training set benign samples), causes its classifier accuracy against
non-adversarial samples to be the lowest of all other defense methods
and the original classifier, making defense sequence GAN unusable
in real-world scenarios. 

Given the low accuracy of the classifier on non-adversarial samples
when using the defense sequence GAN method, as opposed to the good
performance of Defense-GAN in the image recognition domain \cite{DBLP:journals/corr/abs-1805-06605},
we hypothesize that the problem is due to the fact that the sequence
based GANs we used (like SeqGAN) were unable to capture the true distribution
of the input sequence and produced bad seqence GAN output. This was
validated by analyzing the nearest neighbor defense method's performance.\\
We see that for the nearest neighbor defense method, both the classifier
accuracy for non-adversarial test set and the adversarial recall are
higher. This suggests that a larger sequence GAN training set (which
would require a stronger machine with more GPU memory than that used
in this research) would result in a better approximation of the input
sequence distribution by the sequence GAN and thus allow defense sequence
GAN to be as effective as the nearest neighbor defense method - and
more generalizable to samples significantly different from the training
set. Such experiments are planned for future work.\\
The nearest neighbor method also has its limitations: its adversarial
recall against random perturbations is low, since this attack method
adds, on average, more than 50\% new, random API calls. In this case,
the nearest neighbor is more likely to be a benign sample, causing
an adversarial example to be missed.

\section{Conclusions and Future Work}

In this paper, we present a novel defense method against RNN adversarial
examples and compare it to a baseline of four defense methods inspired
by state-of-the-art CNN defense methods. To the best of our knowledge,
this is the first paper to focus on this challenge, which is different
from developing defense methods to use against non-sequence based
adversarial examples. 

We found that sequence squeezing, our novel defense method, is the
only method which provides the balance between the low overhead required
by real time classifiers (like malware classifiers) and good robustness
against adversarial attacks. This makes sequence squeezing the most
suited for real world scenarios, and especially for API call based
RNN classifiers that need to provide classification in real time,
while new classifier input (API calls) is constantly being fed.

Our future work will focus on four directions:
\begin{enumerate}
\item Investigating additional defense methods (e.g., using cluster centroid
distance instead of nearest neighbors to improve run-time performance)
and evaluating the performance obtained when several defense methods
are combined.
\item Extending our work to other domains that use input sequences, such
as NLP.
\item Observing the performance gap of our defense methods against black-box
and white-box attacks (as discussed in Section \ref{subsec:Defense-Methods-Performance})
illustrates some interesting questions. For example, we would evaluate
the effect of implementing defense methods using white-box adversarial
examples, as opposed to black-box examples, which were used in this
paper.
\end{enumerate}
The increasing usage of machine learning based classifiers, such as
``next generation anti virus,'' raises concern from actual adversaries
(the malware developers) trying to evade such systems using adversarial
learning. Making such malware classifiers robust against adversarial
attack is therefore important, and this paper (as well as future research
in this domain) provides methods to defend classifiers that use RNNs,
such as dynamic analysis API call sequence based malware classifiers,
from adversarial attacks.

\newpage{}

\bibliographystyle{plain}
\bibliography{thesis}

\newpage{}

\section*{Appendix A: Tested Dataset}

We used identical implementation details (e.g., dataset, classifier
hyperparameters, etc.) as \cite{DBLP:conf/raid/RosenbergSRE18}, so
the attacks can be compared. Those details are provided here for the
reader's convenience.

\begin{figure*}[tbph]
\begin{centering}
\textsf{\includegraphics[scale=0.75]{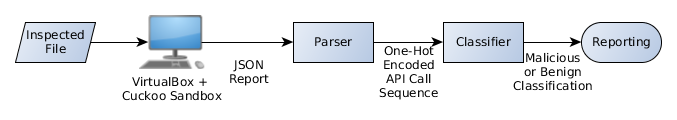}}
\par\end{centering}
\caption{\label{fig:Overview-of-the}Overview of the Malware Classification
Process}
\end{figure*}

An overview of the malware classification process is shown in Figure
\ref{fig:Overview-of-the}, as presented in \cite{DBLP:conf/raid/RosenbergSRE18}.

The dataset used is large and includes the latest malware variants,
such as the Cerber and Locky ransomware families. Each malware type
(ransomware, worms, backdoors, droppers, spyware, PUAs, and viruses)
has the same number of samples, in order to prevent prediction bias
towards the majority class. 20\% of the malware families (such as
the NotPetya ransomware family) were only used in the test set to
assess generalization to an unseen malware family. 80\% of the malware
families (such as the Virut virus family) were distributed between
the training and test sets to determine the classifier's ability to
generalize to samples from the same family. The temporal difference
between the training set and the test set is six months (i.e., all
training set samples are older than the test set samples), based on
VirusTotal's 'first seen' date. 

The ground truth labels of the dataset were determined by \href{https://www.virustotal.com/}{VirusTotal},
an online malware scanning service which contains more than 60 different
security products. A sample with 15 or more positive classifications
(i.e., is considered malicious) from the 60 products is considered
malicious by our classifiers. A sample with zero positive classifications
is labeled as benign. All samples with 1-14 positives were omitted
to prevent false positive contamination of the dataset. Family labels
for dataset balancing were taken from the Kaspersky Anti-Virus classifications.

It is crucial to prevent dataset contamination by malware that detects
whether the malware is running in a Cuckoo Sandbox (or on virtual
machines) and if so, quits immediately to prevent reverse engineering
efforts. In those cases, the sample's label is malicious, but its
behavior recorded in Cuckoo Sandbox (its API call sequence) isn't,
due to its anti-forensic capabilities. To mitigate such contamination
of the dataset, two countermeasures were used: 
\begin{enumerate}
\item Considering only API call sequences with more than 15 API calls (as
in \cite{Huang2016}) and omitting malware that detects a virtual
machine (VM) and quits.
\item Applying \href{https://github.com/Yara-Rules/rules}{YARA rules} to
find samples trying to detect sandbox programs, such as Cuckoo Sandbox,
and omitting all such samples. 
\end{enumerate}
One might argue that the evasive malware that applies such anti-VM
techniques are extremely challenging and relevant, however in this
paper, we focus on the adversarial attack. This attack is generic
enough to work for those evasive malware as well, assuming that other
mitigation techniques (e.g., anti-anti-VM), would be applied. 

After this filtering and balancing of the benign samples, about 400,000
valid samples remained. The final training set size is 360,000 samples,
36,000 of which serve as the validation set. The test set size is
36,000 samples. All sets are balanced between malicious and benign
samples. Due to hardware limitations, a subset of the dataset was
used as a training set: 54,000 training samples and test and validation
sets of 6,000 samples each. The dataset was representative and maintained
the same distribution as the entire dataset described above.

\newpage{}

\section*{Appendix B: Malware Classifiers Tested}

As mentioned in Section \ref{sec:Experimental-Evaluation}, we used
the malware classifiers from \cite{DBLP:conf/raid/RosenbergSRE18},
since many classifiers are covered, allowing us to evaluate the defense
performance against many types of classifiers. The maximum input sequence
length was limited to $m=140$ API calls, since longer sequence lengths,
e.g., $m=1000$, had no effect on the accuracy, and padded shorter
sequences with zeros. A zero stands for a null/dummy value API in
our one-hot encoding. Longer sequences are split into windows of $m$
API calls each, and each window is classified in turn. If any window
is malicious, the entire sequence is considered malicious. Thus, the
input of all of the classifiers is a vector of $m=140$ API call types
in one-hot encoding, using 314 bits, since there were 314 monitored
API call types in the Cuckoo reports for the dataset. The output is
a binary classification: malicious or benign. 

\begin{figure*}[tbph]
\begin{centering}
\textsf{\includegraphics[scale=0.7]{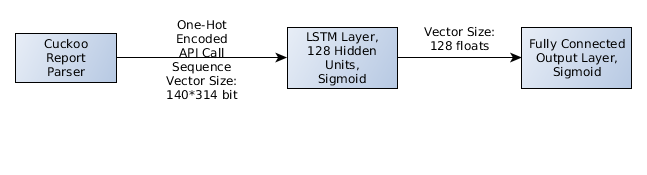}}
\par\end{centering}
\caption{\label{fig:LSTM-Classifier-Architecture}LSTM Classifier Architecture}
\end{figure*}

An overview of the LSTM architecture is shown in Figure \ref{fig:LSTM-Classifier-Architecture}. 

The \href{https://keras.io/}{Keras} implementation was used for all
neural network classifiers, with TensorFlow used for the backend.

The loss function used for training was binary cross-entropy. The
Adam optimizer was used for all of the neural networks. The output
layer was fully connected with sigmoid activation for all neural networks.
For neural networks, a rectified linear unit, $ReLU(x)=max(0,x)$,
was chosen as an activation function for the input and hidden layers
due to its fast convergence compared to $sigmoid()$ or $\tanh()$,
and dropout was used to improve the generalization potential of the
network. A batch size of 32 samples was used.

The classifiers also have the following classifier-specific hyperparameters:
\begin{itemize}
\item RNN, LSTM, GRU, BRNN, BLSTM, bidirectional GRU - a hidden layer of
128 units, with a dropout rate of 0.2 for both inputs and recurrent
states.
\item Deep LSTM and BLSTM - two hidden layers of 128 units, with a dropout
rate of 0.2 for both inputs and recurrent states in both layers.
\end{itemize}
Classifier performance was measured using the accuracy ratio, which
gives equal importance to both false positives and false negatives
(unlike precision or recall). The false positive rate of the classifiers
varied between 0.5-1\%.

The performance of the classifiers is shown in Table \ref{tab:Classifier-Performance}.
The accuracy was measured on the test set, which contains 36,000 samples.

\begin{table}[htbp]
\caption{\label{tab:Classifier-Performance}Classifier Performance}
\centering{}%
\begin{tabular}{|c|c|}
\hline 
Classifier Type & Accuracy (\%)\tabularnewline
\hline 
\hline 
RNN & 97.90\tabularnewline
\hline 
BRNN & 95.58\tabularnewline
\hline 
LSTM & 98.26\tabularnewline
\hline 
Deep LSTM & 97.90\tabularnewline
\hline 
BLSTM & 97.90\tabularnewline
\hline 
Deep BLSTM & 98.02\tabularnewline
\hline 
GRU & 97.32\tabularnewline
\hline 
Bidirectional GRU & 98.04\tabularnewline
\hline 
\end{tabular}
\end{table}

As can be seen in Table \ref{tab:Classifier-Performance}, the LSTM
variants are the best malware classifiers, in terms of accuracy.

\newpage{}

\section*{Appendix C: Adversarial Attacks Used}

We used the adversarial attack implemented in \cite{DBLP:conf/raid/RosenbergSRE18},
since this is the state-of-the-art adversarial attack in the cyber
security domain. An adversarial example is a sequence of API calls
classified as malicious by the classifier, which is perturbed by the
addition of API calls, so that the modified sequence will be misclassified
as benign. In order to prevent damaging the code's functionality,
one cannot remove or modify API calls; one can only add additional
API calls. The attack used to evaluate the robustness of our defense
method is described in Algorithm \ref{alg:Adversarial-Sequence-Generation}.

\begin{algorithm}[tbph]
\caption{\label{alg:Adversarial-Sequence-Generation}Adversarial Sequence Generation}
\begin{centering}
\textbf{Input}: $f$ (black-box model), $\hat{f}$ (substitute model),
\textbf{$\boldsymbol{\mathbf{x}}$} (malicious sequence to perturb,
of length $l$), $n$ (size of adversarial sliding window), $D$ (vocabulary)
\par\end{centering}
\textbf{for} each sliding window $\boldsymbol{\mathbf{w}}_{j}$ of
$n$ API calls in $\boldsymbol{\mathbf{x}}$:

\ \ \ \ $\boldsymbol{\boldsymbol{\mathbf{w}}_{j}}^{*}=\boldsymbol{\boldsymbol{\mathbf{w}}_{j}}$

\ \ \ \ \textbf{while} $f(\boldsymbol{\boldsymbol{\mathbf{w}}_{j}^{*}})=malicious$
and the number of modifications in $\boldsymbol{\boldsymbol{\mathbf{w}}_{j}}^{*}<\left\lceil \frac{2n}{3}\right\rceil $:

\ \ \ \ \ \ \ \ Randomly select an API's position i in $\boldsymbol{\mathbf{w}}$

\ \ \ \ \ \ \ \ \# Insert a new adversarial API in position
$i\in\{1..n\}$:

\ \ \ \ \ $\boldsymbol{\boldsymbol{\mathbf{w}}_{j}^{*}}[i]=\arg\min_{api}||sign(\boldsymbol{\boldsymbol{\mathbf{w}}_{j}}^{*}-\boldsymbol{\boldsymbol{\mathbf{w}}_{j}^{*}}[1:i-1]\perp api\perp\boldsymbol{\boldsymbol{\mathbf{w}}_{j}^{*}}[i:n-1])-sign(J_{\hat{f}}(\boldsymbol{\boldsymbol{\mathbf{w}}_{j}})[f(\boldsymbol{\boldsymbol{\boldsymbol{w}}_{j}})])||$

\ \ \ \ Replace $\boldsymbol{\mathbf{w}}_{j}$ (in $\boldsymbol{\mathbf{x}}$)
with $\boldsymbol{\boldsymbol{\mathbf{w}}_{j}}^{*}$

\textbf{return} (perturbed) $\mathbf{\boldsymbol{x}}$
\end{algorithm}

$D$ is the vocabulary of available features, that is, the API calls
recorded by the classifier. The adversarial API call sequence length
of $l$ might be different than $n$, which is the length of the sliding
window API call sequence that is used by the adversary. Therefore,
like the prediction, the attack is performed sequentially on $\left\lceil \frac{l}{n}\right\rceil $
windows of $n$ API calls. We chose $n=m=140$ (the window size of
the classifier, mentioned in Appendix B). $\perp$ is the concatenation
operation. $\boldsymbol{\boldsymbol{\mathbf{w}}_{j}^{*}}[1:i-1]\perp api\perp\boldsymbol{\boldsymbol{\mathbf{w}}_{j}^{*}}[i:n-1]$
is the insertion of the encoded API vector in position $i$ of $\boldsymbol{\mathbf{w}}_{j}^{*}$.
In each window $\boldsymbol{\mathbf{w}}_{j}$, the adversary randomly
chooses $i$ since he/she does not have any way to better select $i$
without incurring significant statistical overhead. 

Note that an insertion of an API in position $i$ means that the APIs
from position $i..n$ ($\boldsymbol{\mathbf{w}}_{j}^{*}[i:n]$ ) are
``pushed back'' one position to make room for the new API call,
in order to maintain the original sequence and preserve the original
functionality of the code. Since the sliding window has a fixed length,
the last API call, $\boldsymbol{\boldsymbol{\mathbf{w}}_{j}^{*}}[n]$,
is ``pushed out'' and removed from $\boldsymbol{\mathbf{w}}_{j}^{*}$
(this is why the term is $\perp\boldsymbol{\boldsymbol{\mathbf{w}}_{j}^{*}}[i:n-1]$,
as opposed to $\perp\boldsymbol{\boldsymbol{\mathbf{w}}_{j}^{*}}[i:n]$).
The APIs ``pushed out'' from $\boldsymbol{\mathbf{w}}_{j}$ will
become the beginning of $\boldsymbol{\mathbf{w}}_{j+1}$, so no API
is ignored.

The newly added API call is $\boldsymbol{\boldsymbol{\mathbf{w}}_{j}^{*}}[i]=\arg\min_{api}||sign(\boldsymbol{\boldsymbol{\mathbf{w}}_{j}}^{*}-\boldsymbol{\boldsymbol{\mathbf{w}}_{j}^{*}}[0:i]\perp api\perp\boldsymbol{\boldsymbol{\mathbf{w}}_{j}^{*}}[i:n-1])-sign(J_{\hat{f}}(\boldsymbol{\boldsymbol{\mathbf{w}}_{j}})[f(\boldsymbol{\boldsymbol{\boldsymbol{w}}_{j}})])||$.
$sign(J_{\hat{f}}(\boldsymbol{\boldsymbol{\mathbf{w}}_{j}})[f(\boldsymbol{\boldsymbol{\boldsymbol{w}}_{j}})])$
gives us the direction in which we have to perturb the API call sequence
in order to reduce the probability assigned to the malicious class,
$f(\boldsymbol{\mathbf{x}})$, and thus change the predicted label
of the API call sequence. However, the set of legitimate API call
embeddings is finite. Thus, we cannot set the new API to any real
value. We therefore find the API call $api$ in $D$ whose insertion
directs us closest to the direction indicated by the Jacobian as most
impactful on the model\textquoteright s prediction. We iteratively
apply this heuristic until either the classification of the entire
sequence changes (meaning that the classification of each window $\boldsymbol{\mathbf{w}}_{j}$
changes) or until the maximum number of modifications is reached in
a certain window: $\left\lceil \frac{2n}{3}\right\rceil $. In the
latter case, the attack has failed. The use of a maximum number of
allowed perturbations is commonly used in adversarial learning, and
a modification of 66.67\% of the malware API calls (up to 93 adversarial
API calls in each window of $n=140$ API calls) is a very permissive
boundary.

Three variants of this attack were used. In the white-box variant,
we used $\hat{f}(\boldsymbol{\mathbf{x}})=f(\boldsymbol{\mathbf{x}})$,
so the attacker is fully aware of the architecture, weights, and hyperparameters
of the target classifier. In the black-box variant, we used a substitute
model, as used in \cite{DBLP:conf/raid/RosenbergSRE18}. The substitute
model was a GRU with 64 units (different from the malware classifiers
used in Appendix B). Besides the classifier's type and architecture,
a different optimizer was used for the substitute model (ADADELTA
instead of Adam). Finally, in the random perturbation variant, no
substitute model was used, and line 7 was replaced with a randomly
chosen API call type.

\newpage{}

\section*{Appendix D: Sequence Squeezing for API Calls Monitored by Cuckoo
Sandbox}

\begin{table*}[tp]
\caption{Sequence Squeezing for Cuckoo Sandbox Monitored API Calls }

\begin{tabular}{|l||l|}
CertControlStore  & CertControlStore\tabularnewline
CertCreateCertificateContext  & CertCreateCertificateContext\tabularnewline
CertOpenStore  & CertOpenStore\tabularnewline
CertOpenSystemStoreA  & CertOpenStore \tabularnewline
CertOpenSystemStoreW  & CertOpenStore \tabularnewline
CoCreateInstance  & CoCreateInstance\tabularnewline
CoCreateInstanceEx  & CoCreateInstance \tabularnewline
CoGetClassObject  & CoGetClassObject\tabularnewline
CoInitializeEx  & CoInitialize\tabularnewline
CoInitializeSecurity  & CoInitialize \tabularnewline
CoUninitialize  & CoUninitialize\tabularnewline
ControlService  & ControlService\tabularnewline
CopyFileA  & CopyFile\tabularnewline
CopyFileExW  & CopyFile \tabularnewline
CopyFileW  & CopyFile \tabularnewline
CreateActCtxW  & CreateActCtxW\tabularnewline
CreateDirectoryExW  & CreateDirectory\tabularnewline
CreateDirectoryW  & CreateDirectory \tabularnewline
CreateProcessInternalW  & CreateProcess\tabularnewline
CreateRemoteThread  & NtCreateThreadEx\tabularnewline
CreateRemoteThreadEx  & NtCreateThreadEx \tabularnewline
CreateServiceA  & CreateService\tabularnewline
CreateServiceW  & CreateService \tabularnewline
CreateThread  & NtCreateThreadEx \tabularnewline
CreateToolhelp32Snapshot  & CreateToolhelp32Snapshot\tabularnewline
CryptAcquireContextA  & CryptAcquireContext\tabularnewline
CryptAcquireContextW  & CryptAcquireContext \tabularnewline
CryptCreateHash  & CryptCreateHash\tabularnewline
CryptDecodeMessage  & CryptDecrypt\tabularnewline
CryptDecodeObjectEx  & CryptDecrypt \tabularnewline
CryptDecrypt  & CryptDecrypt \tabularnewline
CryptEncrypt  & CryptEncrypt\tabularnewline
CryptExportKey  & CryptKey\tabularnewline
CryptGenKey  & CryptKey \tabularnewline
CryptHashData  & CryptHash\tabularnewline
CryptHashMessage  & CryptHash \tabularnewline
CryptProtectData  & CryptEncrypt \tabularnewline
CryptProtectMemory  & CryptEncrypt \tabularnewline
CryptUnprotectData  & CryptDecrypt \tabularnewline
CryptUnprotectMemory  & CryptDecrypt \tabularnewline
DecryptMessage  & CryptDecrypt \tabularnewline
DeleteFileW  & DeleteFile\tabularnewline
DeleteService  & DeleteService\tabularnewline
DeleteUrlCacheEntryA  & DeleteUrlCacheEntry\tabularnewline
DeleteUrlCacheEntryW  & DeleteUrlCacheEntry \tabularnewline
DeviceIoControl  & DeviceIoControl\tabularnewline
DnsQuery\_A  & DnsQuery\tabularnewline
DnsQuery\_W  & DnsQuery \tabularnewline
DrawTextExA  & DrawTextEx\tabularnewline
DrawTextExW  & DrawTextEx \tabularnewline
EncryptMessage  & CryptEncrypt \tabularnewline
EnumServicesStatusA  & EnumServicesStatus\tabularnewline
EnumServicesStatusW  & EnumServicesStatus \tabularnewline
EnumWindows  & EnumWindows\tabularnewline
ExitWindowsEx  & ExitWindows\tabularnewline
FindFirstFileExA  & FindFirstFileEx\tabularnewline
FindFirstFileExW  & FindFirstFileEx \tabularnewline
FindResourceA  & FindResource\tabularnewline
FindResourceExA  & FindResource \tabularnewline
FindResourceExW  & FindResource \tabularnewline
FindResourceW  & FindResource \tabularnewline
FindWindowA  & FindWindow\tabularnewline
FindWindowExA  & FindWindow \tabularnewline
FindWindowExW  & FindWindow \tabularnewline
FindWindowW  & FindWindow \tabularnewline
GetAdaptersAddresses  & GetAdaptersInfo\tabularnewline
GetAdaptersInfo  & GetAdaptersInfo \tabularnewline
GetAddrInfoW  & getaddrinfo\tabularnewline
GetAsyncKeyState  & GetKeyState\tabularnewline
GetBestInterfaceEx  & GetAdaptersInfo \tabularnewline
GetComputerNameA  & GetComputerName\tabularnewline
GetComputerNameW  & GetComputerName \tabularnewline
GetCursorPos  & GetCursorPos\tabularnewline
GetDiskFreeSpaceExW  & GetDiskFreeSpace\tabularnewline
GetDiskFreeSpaceW  & GetDiskFreeSpace \tabularnewline
GetFileAttributesExW  & GetFileInformation\tabularnewline
GetFileAttributesW  & GetFileInformation \tabularnewline
GetFileInformationByHandle  & GetFileInformation \tabularnewline
GetFileInformationByHandleEx  & GetFileInformation \tabularnewline
GetFileSize  & GetFileInformation \tabularnewline
GetFileSizeEx  & GetFileInformation \tabularnewline
GetFileType  & GetFileInformation \tabularnewline
GetFileVersionInfoExW  & GetFileVersionInfo\tabularnewline
GetFileVersionInfoSizeExW  & GetFileVersionInfo \tabularnewline
GetFileVersionInfoSizeW  & GetFileVersionInfo \tabularnewline
GetFileVersionInfoW  & GetFileVersionInfo \tabularnewline
GetForegroundWindow  & GetForegroundWindow\tabularnewline
GetInterfaceInfo  & GetAdaptersInfo \tabularnewline
GetKeyState  & GetKeyState \tabularnewline
GetKeyboardState  & GetKeyState \tabularnewline
GetNativeSystemInfo  & GetSystemInfo\tabularnewline
GetShortPathNameW  & GetShortPathNameW\tabularnewline
GetSystemDirectoryA  & GetSystemDirectory\tabularnewline
GetSystemDirectoryW  & GetSystemDirectory \tabularnewline
GetSystemInfo  & GetSystemInfo \tabularnewline
GetSystemMetrics  & GetSystemMetrics\tabularnewline
GetSystemTimeAsFileTime  & timeGetTime\tabularnewline
GetSystemWindowsDirectoryA  & GetSystemDirectory \tabularnewline
GetSystemWindowsDirectoryW  & GetSystemDirectory \tabularnewline
GetTempPathW  & GetTempPathW\tabularnewline
GetTimeZoneInformation  & GetTimeZoneInformation\tabularnewline
GetUserNameA  & GetUserName\tabularnewline
GetUserNameExA  & GetUserName \tabularnewline
GetUserNameExW  & GetUserName \tabularnewline
GetUserNameW  & GetUserName \tabularnewline
GetVolumeNameForVolumeMountPointW  & GetVolumePathName\tabularnewline
GetVolumePathNameW  & GetVolumePathName \tabularnewline
GetVolumePathNamesForVolumeNameW  & GetVolumePathName \tabularnewline
GlobalMemoryStatus  & GlobalMemoryStatus\tabularnewline
GlobalMemoryStatusEx  & GlobalMemoryStatus \tabularnewline
HttpOpenRequestA  & InternetConnect\tabularnewline
HttpOpenRequestW  & InternetConnect \tabularnewline
HttpQueryInfoA  & InternetQueryOptionA\tabularnewline
HttpSendRequestA  & InternetOpenUrl\tabularnewline
HttpSendRequestW  & InternetOpenUrl \tabularnewline
IWbemServices\_ExecMethod  & IWbemServices\_Exec\tabularnewline
IWbemServices\_ExecQuery  & IWbemServices\_Exec \tabularnewline
InternetCloseHandle  & InternetCloseHandle\tabularnewline
InternetConnectA  & InternetConnect \tabularnewline
InternetConnectW  & InternetConnect \tabularnewline
InternetCrackUrlA  & InternetCrackUrl\tabularnewline
InternetCrackUrlW  & InternetCrackUrl \tabularnewline
InternetGetConnectedState  & InternetGetConnectedState\tabularnewline
InternetGetConnectedStateExA  & InternetGetConnectedState \tabularnewline
InternetGetConnectedStateExW  & InternetGetConnectedState \tabularnewline
InternetOpenA  & InternetOpen\tabularnewline
InternetOpenUrlA  & InternetOpenUrl \tabularnewline
InternetOpenUrlW  & InternetOpenUrl \tabularnewline
InternetOpenW  & InternetOpen \tabularnewline
InternetQueryOptionA  & InternetQueryOptionA \tabularnewline
InternetReadFile  & InternetReadFile\tabularnewline
InternetSetOptionA  & InternetSetOptionA\tabularnewline
InternetSetStatusCallback  & InternetSetStatusCallback\tabularnewline
InternetWriteFile  & InternetWriteFile\tabularnewline
IsDebuggerPresent  & IsDebuggerPresent\tabularnewline
LdrGetDllHandle  & LdrGetDllHandle\tabularnewline
LdrGetProcedureAddress  & LdrGetProcedureAddress\tabularnewline
LdrLoadDll  & LdrLoadDll\tabularnewline
LdrUnloadDll  & LdrUnloadDll\tabularnewline
LoadResource  & LoadResource\tabularnewline
LoadStringA  & LoadResource \tabularnewline
LoadStringW  & LoadResource \tabularnewline
LookupAccountSidW  & LookupAccountSidW\tabularnewline
LookupPrivilegeValueW  & LookupPrivilegeValueW\tabularnewline
MessageBoxTimeoutA  & MessageBoxTimeout\tabularnewline
MessageBoxTimeoutW  & MessageBoxTimeout \tabularnewline
Module32FirstW  & Module32FirstW\tabularnewline
Module32NextW  & Module32FirstW \tabularnewline
MoveFileWithProgressW  & MoveFileWithProgressW\tabularnewline
NetGetJoinInformation  & NetGetJoinInformation\tabularnewline
NetShareEnum  & NetShareEnum\tabularnewline
NetUserGetInfo  & NetUserGetInfo\tabularnewline
NetUserGetLocalGroups  & NetUserGetInfo \tabularnewline
NotifyBootConfigStatus  & NotifyBootConfigStatus\tabularnewline
NtAllocateVirtualMemory  & NtAllocateVirtualMemory\tabularnewline
NtClose  & NtClose\tabularnewline
NtCreateFile  & NtCreateFile\tabularnewline
NtCreateKey  & NtCreateKey\tabularnewline
NtCreateMutant  & NtCreateMutant\tabularnewline
NtCreateSection  & NtCreateSection\tabularnewline
NtCreateThreadEx  & NtCreateThreadEx \tabularnewline
NtCreateUserProcess  & CreateProcess \tabularnewline
NtDelayExecution  & NtDelayExecution\tabularnewline
NtDeleteFile  & DeleteFile \tabularnewline
NtDeleteKey  & RegDeleteKey\tabularnewline
NtDeleteValueKey  & RegDeleteValue\tabularnewline
NtDeviceIoControlFile  & DeviceIoControl \tabularnewline
NtDuplicateObject  & NtDuplicateObject\tabularnewline
NtEnumerateKey  & RegEnumKey\tabularnewline
NtEnumerateValueKey  & RegEnumValue\tabularnewline
NtFreeVirtualMemory  & NtFreeVirtualMemory\tabularnewline
NtGetContextThread  & NtGetContextThread\tabularnewline
NtLoadDriver  & NtLoadDriver\tabularnewline
NtLoadKey  & RegCreateKey\tabularnewline
NtLoadKeyEx  & RegCreateKey \tabularnewline
NtMapViewOfSection  & NtMapViewOfSection\tabularnewline
NtOpenDirectoryObject  & NtOpenDirectoryObject\tabularnewline
NtOpenFile  & NtOpenFile\tabularnewline
NtOpenKey  & RegOpenKey\tabularnewline
NtOpenKeyEx  & RegOpenKey \tabularnewline
NtOpenMutant  & NtOpenMutant\tabularnewline
NtOpenProcess  & NtOpenProcess\tabularnewline
NtOpenSection  & NtOpenSection\tabularnewline
NtOpenThread  & NtOpenThread\tabularnewline
NtProtectVirtualMemory  & NtProtectVirtualMemory\tabularnewline
NtQueryAttributesFile  & GetFileInformation \tabularnewline
NtQueryDirectoryFile  & NtQueryDirectoryFile\tabularnewline
NtQueryFullAttributesFile  & GetFileInformation \tabularnewline
NtQueryInformationFile  & NtQueryInformationFile\tabularnewline
NtQueryKey  & RegEnumKey \tabularnewline
NtQueryMultipleValueKey  & RegEnumValue \tabularnewline
NtQuerySystemInformation  & GetSystemInfo \tabularnewline
NtQueryValueKey  & RegEnumValue \tabularnewline
NtQueueApcThread  & NtQueueApcThread\tabularnewline
NtReadFile  & NtReadFile\tabularnewline
NtReadVirtualMemory  & NtReadVirtualMemory\tabularnewline
NtRenameKey  & NtRenameKey\tabularnewline
NtResumeThread  & NtResumeThread\tabularnewline
NtSaveKey  & RegCreateKey \tabularnewline
NtSaveKeyEx  & RegCreateKey \tabularnewline
NtSetContextThread  & NtSetContextThread\tabularnewline
NtSetInformationFile  & NtSetInformationFile\tabularnewline
NtSetValueKey  & RegSetValue\tabularnewline
NtShutdownSystem  & shutdown\tabularnewline
NtSuspendThread  & NtSuspendThread\tabularnewline
NtTerminateProcess  & NtTerminateProcess\tabularnewline
NtTerminateThread  & NtTerminateThread\tabularnewline
NtUnloadDriver  & NtUnloadDriver\tabularnewline
NtUnmapViewOfSection  & NtUnmapViewOfSection\tabularnewline
NtWriteFile  & NtWriteFile\tabularnewline
NtWriteVirtualMemory  & NtWriteVirtualMemory\tabularnewline
ObtainUserAgentString  & ObtainUserAgentString\tabularnewline
OleInitialize  & CoInitialize \tabularnewline
OpenSCManagerA  & OpenSCManager\tabularnewline
OpenSCManagerW  & OpenSCManager \tabularnewline
OpenServiceA  & OpenService\tabularnewline
OpenServiceW  & OpenService \tabularnewline
OutputDebugStringA  & OutputDebugStringA\tabularnewline
PRF  & PRF\tabularnewline
Process32FirstW  & Process32FirstW\tabularnewline
Process32NextW  & Process32FirstW \tabularnewline
ReadCabinetState  & ReadCabinetState\tabularnewline
ReadProcessMemory  & ReadProcessMemory\tabularnewline
RegCloseKey  & RegCloseKey\tabularnewline
RegCreateKeyExA  & RegCreateKey \tabularnewline
RegCreateKeyExW  & RegCreateKey \tabularnewline
RegDeleteKeyA  & RegDeleteKey \tabularnewline
RegDeleteKeyW  & RegDeleteKey \tabularnewline
RegDeleteValueA  & RegDeleteValue \tabularnewline
RegDeleteValueW  & RegDeleteValue \tabularnewline
RegEnumKeyExA  & RegEnumKey \tabularnewline
RegEnumKeyExW  & RegEnumKey \tabularnewline
RegEnumKeyW  & RegEnumKey \tabularnewline
RegEnumValueA  & RegEnumValue \tabularnewline
RegEnumValueW  & RegEnumValue \tabularnewline
RegOpenKeyExA  & RegOpenKey \tabularnewline
RegOpenKeyExW  & RegOpenKey \tabularnewline
RegQueryInfoKeyA  & RegQueryInfoKey\tabularnewline
RegQueryInfoKeyW  & RegQueryInfoKey \tabularnewline
RegQueryValueExA  & RegQueryValueEx\tabularnewline
RegQueryValueExW  & RegQueryValueEx \tabularnewline
RegSetValueExA  & RegSetValue \tabularnewline
RegSetValueExW  & RegSetValue \tabularnewline
RegisterHotKey  & RegisterHotKey\tabularnewline
RemoveDirectoryA  & RemoveDirectory\tabularnewline
RemoveDirectoryW  & RemoveDirectory \tabularnewline
RtlAddVectoredContinueHandler  & RtlAddVectoredExceptionHandler\tabularnewline
RtlAddVectoredExceptionHandler  & RtlAddVectoredExceptionHandler \tabularnewline
RtlCompressBuffer  & RtlCompressBuffer\tabularnewline
RtlCreateUserThread  & NtCreateThreadEx \tabularnewline
RtlDecompressBuffer  & RtlDecompress\tabularnewline
RtlDecompressFragment  & RtlDecompress \tabularnewline
RtlRemoveVectoredContinueHandler  & RtlRemoveVectoredExceptionHan\tabularnewline
RtlRemoveVectoredExceptionHandler  & RtlRemoveVectoredExceptionHa\tabularnewline
SHGetFolderPathW  & SHGetFolderPathW\tabularnewline
SHGetSpecialFolderLocation  & SHGetSpecialFolderLocation\tabularnewline
SearchPathW  & SearchPathW\tabularnewline
SendNotifyMessageA  & SendNotifyMessage\tabularnewline
SendNotifyMessageW  & SendNotifyMessage \tabularnewline
SetEndOfFile  & SetEndOfFile\tabularnewline
SetErrorMode  & SetErrorMode\tabularnewline
SetFileAttributesW  & SetFileInformation\tabularnewline
SetFileInformationByHandle  & SetFileInformation \tabularnewline
SetFilePointer  & SetFilePointer\tabularnewline
SetFilePointerEx  & SetFilePointer \tabularnewline
SetFileTime  & SetFileTime\tabularnewline
SetUnhandledExceptionFilter  & SetUnhandledExceptionFilter\tabularnewline
SetWindowsHookExA  & SetWindowsHookEx\tabularnewline
SetWindowsHookExW  & SetWindowsHookEx \tabularnewline
ShellExecuteExW  & system\tabularnewline
SizeofResource  & SizeofResource\tabularnewline
Ssl3GenerateKeyMaterial  & CryptKey \tabularnewline
StartServiceA  & StartService\tabularnewline
StartServiceW  & StartService \tabularnewline
TaskDialog  & TaskDialog\tabularnewline
Thread32First  & Thread32First\tabularnewline
Thread32Next  & Thread32First \tabularnewline
URLDownloadToFileW  & InternetWriteFile \tabularnewline
UnhookWindowsHookEx  & UnhookWindowsHookEx\tabularnewline
UuidCreate  & UuidCreate\tabularnewline
WNetGetProviderNameW  & WNetGetProviderNameW\tabularnewline
WSAConnect  & connect\tabularnewline
WSARecv  & recv\tabularnewline
WSARecvFrom  & recv \tabularnewline
WSASend  & send\tabularnewline
WSASendTo  & send \tabularnewline
WSASocketA  & socket\tabularnewline
WSASocketW  & socket \tabularnewline
WSAStartup  & WSAStartup\tabularnewline
WriteConsoleA  & WriteConsole\tabularnewline
WriteConsoleW  & WriteConsole \tabularnewline
WriteProcessMemory  & WriteProcessMemory\tabularnewline
accept  & accept\tabularnewline
bind  & bind\tabularnewline
closesocket  & closesocket\tabularnewline
connect  & connect \tabularnewline
getaddrinfo  & getaddrinfo \tabularnewline
gethostbyname  & gethostbyname\tabularnewline
getsockname  & getsockname\tabularnewline
ioctlsocket  & setsockopt\tabularnewline
listen  & listen\tabularnewline
recv  & recv \tabularnewline
recvfrom  & recv \tabularnewline
select  & select\tabularnewline
send  & send \tabularnewline
sendto  & send \tabularnewline
setsockopt  & setsockopt \tabularnewline
shutdown  & shutdown \tabularnewline
socket  & socket \tabularnewline
system  & system \tabularnewline
timeGetTime  & timeGetTime \tabularnewline
\end{tabular}
\end{table*}

\end{document}